

\font\titlefont = cmr10 scaled\magstep 4
 2
\font\sectionfont = cmr10
\font\littlefont = cmr5 
\font\eightrm = cmr8

\def\ss{\scriptstyle}
\def\sss{\scriptscriptstyle}

\newcount\tcflag
\tcflag = 0  

\ifnum\tcflag = 0 \magnification = 1200 \fi  

\global\baselineskip = 1.2\baselineskip 
\global\parskip = 4pt plus 0.3pt 
\global\abovedisplayskip = 18pt plus3pt minus9pt
\global\belowdisplayskip = 18pt plus3pt minus9pt
\global\abovedisplayshortskip = 6pt plus3pt
\global\belowdisplayshortskip = 6pt plus3pt

\def\barsoff{\overfullrule=0pt}


\def\endignore{}
\def\ignore #1\endignore{} 

\newcount\dflag
\dflag = 0


\def\monthname{\ifcase\month 
\or January \or February \or March \or April \or May \or June%
\or July \or August \or September \or October \or November %
\or December 
\fi}

\newcount\dummy
\newcount\minute  
\newcount\hour
\newcount\localtime
\newcount\localday
\localtime = \time
\localday = \day

\def\advanceclock#1#2{ 
\dummy = #1
\multiply\dummy by 60
\advance\dummy by #2
\advance\localtime by \dummy
\ifnum\localtime > 1440 
\advance\localtime by -1440
\advance\localday by 1
\fi}

\def\settime{{\dummy = \localtime %
\divide\dummy by 60%
\hour = \dummy 
\minute = \localtime%
\multiply\dummy by 60%
\advance\minute by -\dummy 
\ifnum\minute < 10
\xdef\spacer{0} 
\else \xdef\spacer{}
\fi %
\ifnum\hour < 12
\xdef\ampm{a.m.} 
\else
\xdef\ampm{p.m.} 
\advance\hour by -12 %
\fi %
\ifnum\hour = 0 \hour = 12 \fi 
\xdef\timestring{\number\hour : \spacer \number\minute%
\thinspace \ampm}}}



\def\endtitle{}
\def\title#1\endtitle{\vskip.5in\titlefont
\global\baselineskip = 2\baselineskip 
#1\vskip.4in
\baselineskip = 0.5\baselineskip\rm}

\def\endauthors{}
\def\authors#1\endauthors{#1}

\def\endabstract{}
\def\abstract#1\endabstract{\vskip .3in%
\centerline{\sectionfont\bf Abstract}%
\vskip .1in
\noindent#1}

\def\nopageonenumber{\footline={\ifnum\pageno<2\hfil\else
\hss\tenrm\folio\hss\fi}}  

\newcount\nsection 
\newcount\nsubsection 

\def\section#1{\global\advance\nsection by 1
\nsubsection=0
\bigskip\noindent\centerline{\sectionfont \bf \number\nsection.\ #1}
\bigskip\rm\nobreak}

\def\subsection#1{\global\advance\nsubsection by 1
\bigskip\noindent\sectionfont \sl
\number\nsection.\number\nsubsection)\
#1\bigskip\rm\nobreak}

\def\topic #1{{\medskip\noindent $\bullet$ \it #1:}}
\def\endtopic{\medskip}

\def\appendix#1#2{\bigskip\noindent%
\centerline{\sectionfont \bf Appendix #1.\ #2} 
\bigskip\rm\nobreak} 


\newcount\nref 
\global\nref = 1 

\def\therefs{}


\def\ref#1#2{\xdef #1{[\number\nref]} 
\ifnum\nref = 1\global\xdef\therefs{\item{[\number\nref]} #2\ } %
\else
\global\xdef\oldrefs{\therefs}
\global\xdef\therefs{\oldrefs\vskip.1in\item{[\number\nref]} #2\ }%
\fi%
\global\advance\nref by 1
}

\def\listrefs{\vfill\eject\section{References}\therefs}


\newcount\nfoot 
\global\nfoot = 1 

\def\foot#1#2{\xdef #1{(\number\nfoot)} 
\hskip -0.2cm ${}^{\number\nfoot}$
\footnote{}{\vbox{\baselineskip=10pt
\eightrm \hskip -1cm ${}^{\number\nfoot}$ #2}}
\global\advance\nfoot by 1
}


\newcount\nfig 
\global\nfig = 1
\def\thefigs{} 

\def\figure#1#2{\xdef #1{(\number\nfig)}
\ifnum\nfig = 1\global\xdef\thefigs{\item{(\number\nfig)} #2\ }%
\else
\global\xdef\oldfigs{\thefigs}
\global\xdef\thefigs{\oldfigs\vskip.1in\item{(\number\nfig)} #2\ }%
\fi%
\global\advance\nfig by 1 } 

\def\fig#1{\xdef #1{(\number\nfig)}
\global\advance\nfig by 1 } 


\newcount\ntab
\global\ntab = 1

\def\table#1{\xdef #1{\number\ntab}
\global\advance\ntab by 1 } 


\newcount\cflag
\newcount\nequation
\global\nequation = 1
\def\eqlabel{(1)}

\def\nexteqno{\ifnum\cflag = 0
\global\advance\nequation by 1
\fi
\global\cflag = 0
\xdef\eqlabel{(\number\nequation)}}

\def\lasteqno{\global\advance\nequation by -1
\xdef\eqlabel{(\number\nequation)}}

\def\label#1{\xdef #1{(\number\nequation)}
\ifnum\dflag = 1
{\escapechar = -1
\xdef\draftname{\littlefont\string#1}}
\fi}

\def\clabel#1#2{\xdef\eqlabel{(\number\nequation #2)}
\global\cflag = 1
\xdef #1{\eqlabel}
\ifnum\dflag = 1
{\escapechar = -1
\xdef\draftname{\string#1}}
\fi}

\def\cclabel#1#2{\xdef\eqlabel{#2)}
\global\cflag = 1
\xdef #1{\eqlabel}
\ifnum\dflag = 1
{\escapechar = -1
\xdef\draftname{\string#1}}
\fi}


\def\eeq{}

\def\eqnn #1\eeq{$$ #1 $$}

\def\eq #1\eeq{
\ifnum\dflag = 0
{\xdef\draftname{\ }}
\fi 
$$ #1
\eqno{\eqlabel \rlap{\ \draftname}} $$
\nexteqno}







\def\eqa #1\eeq{
\ifnum\dflag = 0
{\xdef\draftname{\ }}
\fi 
$$ \eqalignno{ #1 } $$
\global\cflag = 0}


\def\ie{{\it i.e.\/}}


\def\anp#1#2#3{{\it Ann.\ Phys. (NY)} {\bf #1} (19#2) #3}

\def\npb#1#2#3{{\it Nucl.\ Phys.} {\bf B#1} (19#2) #3}
\def\plb#1#2#3{{\it Phys.\ Lett.} {\bf #1B} (19#2) #3}

\def\prd#1#2#3{{\it Phys.\ Rev.} {\bf D#1} (19#2) #3}
\def\pr#1#2#3{{\it Phys.\ Rev.} {\bf #1} (19#2) #3}

\def\prl#1#2#3{{\it Phys.\ Rev.\ Lett.} {\bf #1} (19#2) #3}


\global\nulldelimiterspace = 0pt



\def\frac#1#2{{{#1} \over {#2}}\,}  
\def\hf{{1\over 2}}
\def\nth#1{{1\over #1}}


\def\Asl{\hbox{/\kern-.7500em\it A}} 
\def\Dsl{\hbox{/\kern-.6700em\it D}} 
\def\dsl{\hbox{/\kern-.5300em$\partial$}}
\def\pxpsl{\hbox{/\kern-.5600em$p$}}
\def\sslsh{\hbox{/\kern-.5300em$s$}}
\def\epssl{\hbox{/\kern-.5100em$\epsilon$}}
\def\delsl{\hbox{/\kern-.6300em$\nabla$}}
\def\lxpsl{\hbox{/\kern-.4300em$l$}}
\def\elxpsl{\hbox{/\kern-.4500em$\ell$}}
\def\kxpsl{\hbox{/\kern-.5100em$k$}}
\def\qxpsl{\hbox{/\kern-.5000em$q$}}
\def\sla#1{\raise.15ex\hbox{$/$}\kern-.57em #1}



\def\twi{\widetilde}

\def\roughly#1{\mathrel{\raise.3ex
\hbox{$#1$\kern-.75em\lower1ex\hbox{
$\sim$}}}}

\def\ol#1{\overline{#1}}





\def\Sca{{\cal A}}
\def\Scb{{\cal B}}

\def\Scd{{\cal D}}

\def\Scf{{\cal F}}

\def\Scl{{\cal L}}
\def\Scm{{\cal M}}

\def\Scq{{\cal Q}}
\def\Scr{{\cal R}}
\def\Scs{{\cal S}}

\def\Scw{{\cal W}}


\def\ssa{{\sss A}}
\def\ssb{{\sss B}}

\def\ssf{{\sss F}}

\def\ssr{{\sss R}}

\def\ssx{{\sss X}}


\def\pmb#1{\setbox0=\hbox{#1}%
\kern-.025em\copy0\kern-\wd0
\kern.05em\copy0\kern-\wd0
\kern-.025em\raise.0433em\box0}


\font\jlgtenbrm=cmbx10
\font\jlgtenbit=cmmib10
\font\jlgtenbsy=cmbsy10
\font\jlgsevenbrm=cmbx10 at 7pt
\font\jlgsevenbsy=cmbsy10 at 7pt
\font\jlgsevenbit=cmmib10 at 7pt
\font\jlgfivebrm=cmbx10 at 5pt
\font\jlgfivebsy=cmbsy10 at 5pt
\font\jlgfivebit=cmmib10 at 5pt
\newfam\jlgbrm

\textfont\jlgbrm=\jlgtenbrm
\scriptfont\jlgbrm=\jlgsevenbrm
\scriptscriptfont\jlgbrm=\jlgfivebrm
\newfam\jlgbit

\textfont\jlgbit=\jlgtenbit
\scriptfont\jlgbit=\jlgsevenbit
\scriptscriptfont\jlgbit=\jlgfivebit
\newfam\jlgbsy

\textfont\jlgbsy=\jlgtenbsy
\scriptfont\jlgbsy=\jlgsevenbsy
\scriptscriptfont\jlgbsy=\jlgfivebsy
\newcount\jlgcode
\newcount\jlgfam
\newcount\jlgchar
\newcount\jlgtmp
\def\bolded#1{
        \jlgcode\the#1 \divide\jlgcode by 4096
        \jlgtmp\the\jlgcode \multiply\jlgtmp by 4096
        \jlgfam\the#1 \advance\jlgfam by -\the\jlgtmp
        \divide\jlgfam by 256
        \jlgtmp\the\jlgcode \multiply\jlgtmp by 16
        \advance\jlgtmp by \the\jlgfam
        \multiply\jlgtmp by 256
        \jlgchar\the#1 \advance\jlgchar by -\the\jlgtmp
        \advance\jlgfam by \the\jlgbrm
        \jlgtmp\the\jlgcode
        \multiply\jlgtmp by 16
        \advance\jlgtmp by \the\jlgfam
        \multiply\jlgtmp by 256
        \advance\jlgtmp by \the\jlgchar
        \mathchar\the\jlgtmp
}


\def\Tr{\mathop{\rm Tr}}
\def\det{\mathop{\rm det}}

\def\Re{{\rm Re\;}}
\def\Im{{\rm Im\;}}



\def\Avg#1{\left\langle #1 \right\rangle}






\nopageonenumber
\baselineskip = 18pt
\barsoff


\def\Fterm#1{\Bigl( #1 \Bigr)_\ssf}
\def\bk{\item{}}

\def\Uonebar{{\ol{U}_1}}
\def\Utwobar{{\ol{U}_2}}
\def\Nonebar{{\ol{N}_1}}
\def\Ntwobar{{\ol{N}_2}}

\def\Dflat{$D^\flat$}

\def\a{{\alpha}}
\def\b{{\beta}}
\def\L{{\Lambda}}

\def\Ggp{$SU(N_1) \times SU(N_2)$}


\line{hep-th/9808087  \hfil McGill-98/16 ,
IFUNAM FT98-9.}

\title
\centerline{Supersymmetric Models with Product Groups }
\centerline{and Field Dependent Gauge Couplings}
\endtitle

\vskip 0.1in
\authors
\centerline{C.P. Burgess${}^a$, A. de la Macorra${}^b$,
I. Maksymyk${}^c$ and F. Quevedo${}^b$}
\vskip .05in
\centerline{\it ${}^a$ Physics Department, McGill University}
\centerline{\it 3600 University St., Montr\'eal, Qu\'ebec, Canada,
H3A 2T8.}
\vskip .05in
\centerline{\it ${}^b$ Instituto de F\'isica, Universidad Nacional
Aut\'onoma de M\'exico}
\centerline{\it Apartado Postal 20-364, 01000  M\'exico D.F.,
M\'exico.}
\vskip .05in
\centerline{\it ${}^c$ TRIUMF, 4004 Wesbrook Mall}
\centerline{\it Vancouver, British Columbia, Canada, V6T 2A3.}
\endauthors

\abstract
\vbox{\baselineskip 15pt
We study the dilaton-dependence of the
effective action for $N=1$,  $SU(N_1) \times SU(N_2)$ models with one
generation of vectorlike matter transforming in the
fundamental of both groups. We treat in detail the confining
 and Coulomb phases of these models writing explicit expressions
 in many cases for the effective superpotential. We can do so for the
 Wilson superpotentials of the Coulomb phase when $N_2=2$, $N_1=2,4$.
 In these cases the Coulomb phase involves a single $U(1)$ gauge
 multiplet,  for which we exhibit the gauge coupling in terms of the modulus of
an
 elliptic curve.  The $SU(4) \times SU(2)$ model reproduces the weak-coupling
 limits in a nontrivial way. In the confining phase of all of these models,
the dilaton superpotential has a runaway form, but in the Coulomb phase
 the dilaton enjoys flat directions.  Had we used the standard  moduli-space
variables:
 $\Tr \Scm^k$, $k=1, \cdots , N_2$, with $\Scm$ the quark condensate matrix,
 to parameterize the flat directions instead of the eigenvalues of $\Scm$,
  we would find physically unacceptable behaviour,  illustrating the importance
 to correctly identify the moduli.}
\endabstract


\vfill\eject

\section{Introduction}

\ref\review{For reviews, see D. Amati, K. Konishi,
Y. Meurice, G.C. Rossi and G. Veneziano, {\it Phys. Rep.} 162 (1989) 169;\bk
K.Intriligator and N.Seiberg, {\it Nucl. Phys.}{\bf B}(Proc. Suppl.)
{\bf 45B,C} (1996) 1;\bk
M.E.Peskin,  in the proceedings of TASI 96, {\it Fields,
Strings, and Duality}, {\tt hep-th/9702094};\bk
M.Shifman, preprint TPI-MINN-97/09-T, {\tt hep-th/9704114}.}

\ref\adsone{I.Affleck, M.Dine, N.Seiberg, \npb{241}{84}{493}}
\ref\adstwo{I.Affleck, M.Dine, N.Seiberg, \npb{256}{85}{557}}
\ref\seiberg{N.Seiberg, \prd{49}{94}{6857}}

Great strides have been recently made
in the understanding of
nonperturbative effects in supersymmetric
field theories \review\ \adsone\ \adstwo\ \seiberg:
Seiberg and other workers have
developed methods that allow us
to write the exact
form of the low-energy superpotential for
many supersymmetric gauge theories.  Using
these methods, we explore the vacua and low-energy
limit in a class of $N=1$ supersymmetric
gauge theories with gauge
group $G = SU(N_1) \times SU(N_2)$.

\ref\pst{E.Poppitz, Y.Shadmi and S.P.Trevedi, \npb{480}{96}{125}}
\ref\lestringmodels{For a review see for instance, F. Quevedo,
 {\tt hep-th/9603074}.}

We start by adding our motivations for the
study of supersymmetric gauge theories with product
gauge groups to those already given in ref.~\pst.
The strongest reason for their study is based on
the following two observations. First, product
gauge groups are ubiquitous in `realistic' applications,
including low-energy string models \lestringmodels.
Second, their low-energy properties turn out to be
interestingly different from those of supersymmetric
theories involving simple gauge groups.

A feature of these models which is of particular
interest to us is the dilaton dynamics which
they predict. That is, in string theory
the dilaton couples to low-energy
gauge bosons in the following way:
\label\sdef
\eq
\Scl_{\rm kin} =  \nth4 \sum_{r=1}^2
\Fterm{ S_r \, \Tr \Scw_r \Scw_r} ,
\eeq
where $\Fterm{\cdots}$ denotes the chiral supersymmetric
invariant, $\Scw_r$ is the gauge-kinetic chiral
spinor supermultiplet, and the $S_r$'s are related
to the dilaton superfield, $S$, by the well-known relation
\label\srdefs
\eq
S_r = k_r \, S .
\eeq
Here the $k_r$ are the Kac-Moody levels
of the corresponding gauge-group factors.
Moduli dependence due to threshold
effects, or extra contributions to $S_r$ due to nonperturbative string
dynamics can also be considered.

The low-energy scalar potential
for the dilaton is important because
eq.~\sdef\ implies the vacuum values of the scalar
components, $s_r$, of the superfields $S_r$,
play the role of the low-energy gauge coupling, $g_r$,
being related to these couplings and the vacuum
angles, $\Theta_r$, by:
\label\gbaredef
\eq
s_r = { 1 \over g^2_{r}} -
{i \Theta_{r} \over 8 \pi^2}.
\eeq
Moreover, within string theory general nonrenormalization
theorems imply that the dilaton scalar
potential is not generated in perturbation theory,
and so its determination is a strong-coupling problem.

\ref\racetrack{N.V.Krasnikov, \plb{193}{87}{370}}

\ref\us{C.P.Burgess, F.Quevedo, A.de la Macorra, I.Maksymyk,
\plb{410}{97}{181}, {\tt hep-th/9707062}.}

Some generic, and troubling, features of the low-energy dilaton
superpotential, $W(S)$, in four-dimensional supersymmetric
string vacua have emerged over the past decade
and a half of study. After elimination of other
low-energy fields, these superpotentials
typically have the form \racetrack:
\label\genformforW
\eq
W(S) = \sum_k A_k e^{- a_k \, S},
\eeq
where $A_k$ and $a_k$ are numbers which depend
on the model considered, with $a_k$ generally
positive.\foot\divexp{A class of models for which some of
the $\ss a_k$ are negative have recently been
constructed in ref.~\us, using nonabelian
gauge groups which are not asymptotically-free.}
Such a superpotential implies a scalar potential
with a runaway minimum, at $\Re s \to \infty$.

\ref\genericprob{M. Dine and N. Seiberg, \plb{162}{85}{299}.}

In fact, the existence of this kind of
runaway solution appears to be
model independent, as may be seen from
eq.~\gbaredef. Since the dilaton
plays the role of the gauge coupling, and
since flat space with zero coupling is a
well-known string vacuum, the scalar potential
for the low-energy modes of four-dimensional
string solutions generally
lead to scalar potentials
for which the dilaton is driven towards
zero coupling: $\Re s \to \infty$ \genericprob.
Such runaway behaviour follows quite
generally so long as these models are continuous
in the zero-coupling limit. (It is this last continuity
assumption which is evaded in the models
of ref.~\us.)

The low-energy dilaton dynamics which we find for
the product groups explored here is as follows.
The models exhibit several phases, and the low-energy
degrees of freedom which arise depend on which phase
is involved. The models typically exhibit a
confining phase, for which the nonabelian
gauge dynamics are confined, with a
gap between the strongly-coupled ground state
and its low-energy excitations. In this phase
we find eq.~\genformforW\ applies, leading
to the usual dilaton runaway.

There is also
another, Coulomb, phase which involves
massless degrees of freedom, and in this phase
we find a low-energy (Wilson) superpotential
having flat directions for the dilaton field,
even after the strong-coupling effects are included.
This phase therefore differs from eq.~\genformforW\
inasmuch as the dilaton is not {\it driven} to
infinity by strong-interaction effects. What value
its v.e.v. ultimately takes cannot be determined
without more information, in particular as to
how supersymmetry is ultimately spontaneously broken.

\ref\ccm{B.de Carlos, J.A.Casas and C.Mu\~noz, \npb{399}{93}{623}}
\ref\wu{Yi-Yen Wu, Doctoral Thesis, UCB Report No. UCB-PTH-97/18,
     {\tt hep-th/9706040}.}
\ref\intr{K.Intriligator, S.Thomas, \npb{473}{96}{121}.}
\ref\exampleone{ C.Cs\'aka, L.Randall and W.Skiba, \npb{479}{96}{65}}
\ref\exampletwo{A.Nelson, \plb{369}{96}{277}}
\ref\ck{P.Cho and P.Kraus, \prd{54}{96}{7640}}
\ref\SeIn{K. Intriligator and N. Seiberg, \npb{431}{94}{551},
{\tt hep-th/9408155}.}

The models we consider have gauge group $G = SU(N_1)
\times SU(N_2)$, with $N_1 \ge N_2 \ge 2$.
We choose matter  which transforms only in the fundamental
representation of both of the factors of the gauge group,
$R = ({\bf N_1}, {\bf N_2}) \oplus ({\bf \Nonebar},{\bf \Ntwobar})$.
Although the inclusion of field-dependent gauge
couplings for product gauge groups
is not novel in itself, earlier workers have
not done so for matter field carrying charges for
more than one factor of  a product gauge group
\ccm\ \wu. Models having the gauge group
$G = SU(N_1) \times SU(N_2)$ have
also been examined by other authors  \adstwo\ \pst\
\intr\ \exampleone\ \exampletwo\ \ck, although with
a matter content which differs from what we consider here.
The special case where $N_1 = N_2 = 2$ is also
analyzed in ref.~\SeIn.

We present our results in the following way.
\S2\ starts with the construction of the effective
superpotential for the factor-group models we wish to
explore.  After presenting some preliminaries,
we state the general symmetries and limiting
behaviour which guide the determination of
the model's effective superpotential. Because
their low-energy behaviour differs dramatically,
we consider separately the cases where the
mass of the quark supermultiplets is zero (the
Coulomb phase) and nonzero (the
confining phase).

We explore our first application of the general
results in \S3, where we solve in explicit detail
for the superpotential and gauge coupling function
of a simple illustrative model, consisting
of one generation of matter transforming as a
$({\bf 2}, {\bf 2}) \oplus ( {\bf \ol{2}}, {\bf \ol{2}})$
of the gauge group $SU(2) \times SU(2)$. We argue
that the Coulomb phase of this model has a low-energy
superpotential which is completely dilaton-independent,
evading the problem of the runaway dilaton by making
the dilaton a {\it bona fide} flat direction, even
after the inclusion of nonperturbative quantum effects.
Our results for this particular model reproduce those of ref.~\SeIn.

Next, \S4\ presents another simple model, consisting
of one generation of matter transforming as a
$({\bf 4}, {\bf 2}) \oplus ( {\bf \ol{4}}, {\bf \ol{2}})$
of the gauge group $SU(4) \times SU(2)$. The weak-coupling
limit of this model has interesting complications
because its low-energy spectrum changes fundamentally
in the limit of vanishing $SU(4)$ coupling. We
are led in a different way to a similar conclusion
as for the model of \S3: to a low-energy superpotential
with directions along which the dilaton can vary
without breaking supersymmetry, and so
with no cost in energy. We also present ans\"atze
for the gauge-coupling functions for the Coulomb phases
of the models of \S3\ and \S4.

Finally, \S5\ briefly summarizes our conclusions.

\section{$SU(N_1) \times SU(N_2)$ Models}

\ref\vy{G.Veneziano, S.Yankielowicz, \plb{113}{82}{231}}
\ref\bdqq{C.P. Burgess, J.P. Derendinger, F. Quevedo, M. Quir\'os,
\anp{250}{96}{193}.}

We now collect results which apply generally to
models having gauge group \Ggp, with
one generation of matter fields:
\label\matreps
\eq
Q_{a\a} \in ({\bf N_1}, {\bf N_2}) \qquad
\hbox{and} \qquad \twi Q^{a\a} \in
({\bf \ol{N}_1}, {\bf \ol{N}_2}).
\eeq
We use here $a,b,c,\dots$
as the gauge indices of $SU(N_1)$, and $\a,\b,\gamma,\dots$
as those of $SU(N_2)$.
We may take, without loss of generality,
$N_1 \ge N_2 \ge 2$. Finally, we assume the microscopic superpotential
to involve only a quark mass term:
\label\pqmass
\eq
w(Q, \twi Q) = m \; Q_{a\a} \twi Q^{a\a}.
\eeq
Except for the special case $N_1 = N_2 = 3$,
this is the only term possible which is both renormalizable
and gauge invariant.

Our goal is to construct the effective superpotential
and, where relevant, the effective gauge couplings of
this model. We do so following what have become
standard methods, and those readers interested in
the applications to the $SU(2) \times SU(2)$ and
$SU(4) \times SU(2)$ models can proceed directly
to \S3\ and \S4. Our notation and the details
of our procedure follow those
of ref.~\bdqq, (see also ~\vy\ ) and we also temporarily maintain
the fiction that the quantities $S_r$ are independent
fields, with the connection to the single dilaton field, $S$,
through eq.~\srdefs, deferred to the
final expressions.

\subsection{The Semiclassical Spectrum}

We start by sketching the low-energy phases which
are indicated semiclassically, directly using
the microscopic degrees of freedom.
These are determined by examining
the minima of the classical scalar potential,
\label\scapot
\eq
V = \Bigl| \Scf_{a\a} \Bigr|^2 +
\Bigl| \twi\Scf^{a\a} \Bigr|^2 +
\hf \; D_\ssa^2.
\eeq
where $\Scf_{a\a} = \Bigl( \partial w/\partial \Scq_{a\a} \Bigr)^*$,
$\twi\Scf^{a\a} = \Bigl( \partial w/\partial \twi\Scq^{a\a} \Bigr)^*$,
and $D_\ssa = \Scq^\dagger T_\ssa \Scq + \twi{\Scq}^\dagger
\twi T_\ssa \twi\Scq$. Here $\Scq_{a\a}$
and $\twi{\Scq}^{a\a}$ represent the scalar components of the
superfields $Q_{a\a}$ and $\twi{Q}^{a\a}$, while
$T_\ssa$ and $\twi T_\ssa$ represent the generators
of the gauge group acting on these fields. Finally,
$w$ represents the microscopic superpotential, given
by eq.~\pqmass.

Clearly the scalar potential differs
qualitatively according to whether
or not the quark masses satisfy $m = 0$,
since if this is true the superpotential $w$
identically vanishes. We therefore present the
semiclassical analysis separately
for these two cases.

\topic{The Coulomb Phase ($m = 0$)}

Consider first the case $m=0$, for which the
microscopic superpotential vanishes.
In this case the classical scalar potential
vanishes along any scalar field configuration
for which the $D_\ssa$ vanish.
These conditions define the following $D$-flat
($D^\flat$) directions:
\label\Dflatdirns
\eq
\Scq_{a\a} = \pmatrix{v_1 & &  \cr
& \ddots & \cr
& & v_{N_2} \cr
0 & \cdots & 0\cr
\vdots && \vdots \cr
0 & \cdots & 0\cr}
\qquad  \hbox{and} \qquad
\twi \Scq^{a\a} = \pmatrix{\tilde v_1 & &  \cr
& \ddots & \cr
& & \tilde v_{N_2} \cr
0 & \cdots & 0\cr
\vdots && \vdots \cr
0 & \cdots & 0\cr} ,
\eeq
provided the nonzero coefficients satisfy $v_i = \tilde v_i^*$, for
each $i = 1,\dots,N_2$. These field configurations
do not break supersymmetry, but
for generic values the gauge group is broken
down to a subgroup, $H$. Three cases
arise, each having a different $H$:

\item{1.}
If $N_1 = N_2$ the unbroken subgroup is simply
$H= [U(1)]^{N_2 - 1}$, where each of the
$U(1)$ factors is in a diagonal subgroup
of the two gauge-group factors.

\item{2.}
If $N_1 = N_2 + 1$ the unbroken subgroup becomes
$H= [U(1)]^{N_2}$, where the additional
$U(1)$ factor corresponds to phase rotations
of the bottom row of $\Scq_{a\a}$ and
$\twi{\Scq}_{a\a}$.

\item{3.}
If $N_1 \ge N_2 + 2$ the unbroken subgroup is
$H= SU(N_1 - N_2) \times
[U(1)]^{N_2 - 1}$.

For special values of $v_i$ and $\tilde v_i$ the
unbroken semiclassical symmetry group can be larger than this.

We are therefore led, semiclassically for $m = 0$,
to a Coulomb phase in which the low-energy theory
is supersymmetric, containing several $U(1)$ gauge multiplets,
plus a number of matter multiplets which parameterize
the potential's $D^\flat$ directions.\foot\poppitz{We thank Eric Poppitz
for identifying an  error in our previous treatment
of the $\ss D^\flat$ directions.} (Any nonabelian factors
of the gauge group are expected to confine, and so to
drop out of the very-low-energy sector.)

\ref\gaugeinvariants{F. Buccella, J.P. Derendinger,
S. Ferrara and C.A. Savoy, \plb{115}{82}{375};\bk
M.A. Luty and W. Taylor IV, \prd{53}{96}{3399}.}

\ref\pr{E. Poppitz and L. Randall, \plb{336}{94}{402}.}

The number of complex fields required to
parameterize these semiclassical flat directions
\gaugeinvariants, \pr\
is the total number of complex scalar fields, $\Scs = 2 N_1 N_2$,
less the number of broken generators of the gauge group,
$\Scb = \hbox{dim} \; (G/H)$. The superpotential of the
Wilson action for these low-energy degrees of
freedom therefore requires $\Scd = \Scs - \Scb$ matter fields
as its arguments, describing these \Dflat\ directions.

For each of the three
cases for $H$ considered above we therefore find:

\item{1.}
If $N_1 = N_2$ then $H= [U(1)]^{N_2 - 1}$, so
the semiclassical low-energy spectrum contains
$(N_2 -1)$ $U(1)$ gauge supermultiplets and
$\Scd = 2 N_2^2 - [2 (N_2^2 - 1) - (N_2 -1)]
= N_2 + 1$ \Dflat\ directions.

\item{2.}
If $N_1 = N_2 + 1$ then $H= [U(1)]^{N_2}$, so
the semiclassical low-energy spectrum contains
$N_2$ $U(1)$ gauge multiplets and
$\Scd = 2 (N_2+1) N_2 - \left[ \Bigl((N_2+1)^2 - 1
\Bigr) + (N_2^2 - 1) - N_2 \right] = N_2 + 1$ \Dflat\ directions.

\item{3.}
If $N_1 \ge N_2 + 2$ then $H= SU(N_1 - N_2) \times
[U(1)]^{N_2 - 1}$, so
the semiclassical low-energy spectrum contains
$(N_2 -1)$ $U(1)$ gauge supermultiplets and
$\Scd = 2 N_1 N_2 - \left[(N_1^2 -1) + (N_2^2 - 1)
- \Bigl( (N_1 - N_2)^2 - 1 \Bigr) - (N_2-1)
\right] = N_2$
\Dflat\ directions. As mentioned earlier, the nonabelian
$SU(N_1 - N_2)$ gauge multiplet
is expected to confine and so to decouple from
the low-energy theory.

\topic{The Confining Phase ($m \ne 0$)}
If $m \ne 0$, then the degeneracy along the $D^\flat$
directions is directly lifted, even semiclassically,
by the microscopic superpotential, eq.~\pqmass, indicating
that the squark fields vanish in the vacuum.
In this case the semiclassical massless spectrum
therefore consists of a nonabelian $SU(N_1) \times
SU(N_2)$ supersymmetric gauge multiplet, with no
massless matter multiplets. Keeping in mind that
the gauge multiplets are expected to confine,
we therefore expect in this case a gapped low-energy
theory with no massless states.
\endtopic

In the following sections we explore these two phases in
considerably more detail.

\subsection{Which Effective Superpotential?}

There are two kinds of superpotentials which are useful
for exploring the vacuum and low-energy
properties of these (and other) supersymmetric gauge
theories. On the one hand there is the superpotential
for the `exact quantum effective action', which
generates the irreducible correlation functions of
the theory. The arguments of this superpotential can be
chosen to be any fields whose correlations are to
be studied. The other superpotential is that for
the `Wilson' action which describes the dynamics of
the theory's low-energy modes.

For the present
purposes the following properties of these superpotentials
are the most important:\foot\npivswilson{See ref. \bdqq\ for
more details concerning the definitions and differences between
these superpotentials within the context of supersymmetric
gauge theories.}
\topic{(1) Locality}
Because the Wilson action receives no contributions from
massless states, it is guaranteed to be a local quantity.
This property is crucial, since it underlies the holomorphy
of the superpotential which determines the vacuum properties
\review.
The same need {\it not} be true for the quantum action
if the system involves massless degrees of freedom.
\topic{(2) Linearity}
As is proven in \bdqq, if the arguments of the
quantum action are taken to include
the variables
\label\conddef
\eq
{M} \equiv \Avg{ Q_{a\a} \tilde{Q}^{a\a}}, \qquad
\hbox{and}\qquad U_r \equiv \Avg{ \Tr \Scw_r \Scw_r } ,
\quad r = 1,2
\eeq
then the definitions imply the conjugate quantities,
$m$ and $S_r$, can appear in the superpotential
only through the
terms $m M$ and $\nth4 \sum_r S_r U_r$.
\topic{(3) Equivalence}
For systems having no massless degrees of freedom,
it can happen that the quantum action coincides with
the Wilson action if their arguments are chosen to
be the same fields. This is because both are local
due to the absence of massless states, and the
symmetries of the problem may then uniquely
determine the form of the result.
\endtopic

Which of these actions is relevant depends on
the question of physical interest, and on which
of the theory's phases is under consideration. For example,
for the Coulomb phase ($m=0$) it is the effective
superpotential and gauge-coupling function for
`the' Wilson action governing the dynamics of the
massless modes which we construct. (The word `the'
appears in quotations here because in reality there
is potentially a different Wilson action for each
vacuum of the model.) For the confining phase, on
the other hand, it is the superpotential for both
the Wilson and quantum effective actions which we
compute. Both are local because of the absence
of gapless modes, and the symmetries of the model
force them to be identical when evaluated for
appropriately chosen field configurations.

\subsection{Global Symmetries}

In order to determine the form of the superpotential
of the effective theories we take advantage of
the global symmetry group which the model enjoys when $m=0$.
(This would also be a symmetry when $m \ne 0$, provided
we also transform $m$ appropriately.) For generic
values of $N_1$ and $N_2$, this symmetry group is
$U_\ssa(1) \times U_\ssb(1) \times U_\ssr(1)$,
defined by:
\label\produltransfos
\eq
\eqalign{
Q_{a\a}(\theta)  &\rightarrow e^{ i \beta_\ssa + i \beta_\ssb + 2i
\beta_\ssr/3}
Q_{a\a}\left(e^{i\beta_\ssr}\theta\right)  , \cr
\twi{Q}^{a\a}(\theta)  &\rightarrow  e^{ i \beta_\ssa -
i \beta_\ssb + 2i \beta_\ssr/3}
\twi{Q}^{a\a}\left(e^{i\beta_\ssr}\theta\right)  , \cr
\Scw_r(\theta)  &\to  e^{i \beta_\ssr} \Scw_r
\left(e^{i\beta_\ssr}\theta\right), \cr}
\eeq
where $\beta_\ssa$,  $\beta_\ssb$ and $\beta_\ssr$
are the transformation parameters.

The effective superpotential is constructed by requiring
it to realize these symmetries in the same way as does
the microscopic theory. For instance, if the arguments
of the superpotential are the variables
$M$ and $U_r$ of eq.~\conddef, then the action of
the global symmetry follows from
eqs.~\conddef\ and \produltransfos:
\label\transfos
\eq
M(\theta)  \rightarrow  e^{ 2 i \beta_\ssa + 4 i \beta_\ssr/3 }
M \left(e^{i\beta_\ssr}\theta\right)\qquad
U_r(\theta)  \rightarrow  e^{ 2 i \beta_\ssr } U_r
\left(e^{i\beta_\ssr}\theta\right) .
\eeq
The anomaly-free symmetry, $U_\ssb(1)$, must simply be a
symmetry of $W(U_r,S_r,M)$. For the anomalous $U_\ssa(1)
\times U_\ssr(1)$ transformations,
however, $W$ must reflect the
microscopic theory's property that shifts of
the $S_r$ are required to cancel the anomalies.
Since the various symmetries have separate anomalies with
each of the gauge group factors --- all
mixed anomalies vanish which involve both gauge groups
simultaneously --- independent shifts are
required for each of the superfields $S_r$.
These are possible so long as these fields are regarded
as being independent of one another.

The upshot is that the effective action must be invariant
with respect to the anomalous symmetries, provided that eq.~\transfos\
is supplemented by an appropriate transformation law for $S_r$.
The required transformation is simply formulated in terms of the
fields $L_r \equiv \exp[- 4 \pi^2 S_r]$, for which:
\label\dshiftprodt
\eq
L_r(\theta)  \rightarrow e^{i \Sca^r_\ssa \beta_\ssa
+ i \Sca^r_\ssr \beta_\ssr } \;
L_r\left(e^{i\beta_\ssr} \theta\right) ,
 \qquad
\hbox{with} \qquad \Sca^r_\ssx =   \sum_i T(\Scr^r_i) Q_\ssx(\Scr^r_i) .
\eeq
Here $X = A,R$ distinguishes the two anomalous symmetries,
and $\sum_i$ is a sum over the gauge representations, $\Scr^r_i$, of
all of the left-handed spin-half fields of the model.
$Q_\ssx(\Scr^r_i)$ denotes the quantum
number of these fields under the anomalous
symmetry $X = A,R$. $T(\Scr^r_i)$ is defined
in terms of the trace of the gauge generators in the representation
of interest, {\it via} $\Tr_{\Scr^r} [ t_a t_b ]
\equiv T(\Scr^r) \; \delta_{ab}$.   We use the standard convention
for which the gauge generators are normalized so that
$T(F) = {1\over 2}$ in the fundamental representation, and so
then $T(A) = N_c$ for the adjoint representation of $SU(N_c)$.
For instance, when evaluated
for supersymmetric QCD (SQCD) with $N_f$ quark flavours,
the quarks give $\sum_i T(\Scr_i) = N_f$, while for the gauginos we have
$T(A) = N_c$.

For the product model of interest the coefficients $\Sca^r_\ssx$,
appearing in eq.~\dshiftprodt, become:
\label\dshiftanoms
\eq
\eqalign{
\Sca^1_\ssa = {N_2},  \qquad&\qquad
\Sca^2_\ssa = {N_1},  \cr
\Sca^1_\ssr =  N_1 - {N_2 \over 3},  \qquad&\qquad
\Sca^2_\ssr =  N_2 - {N_1 \over 3} . \cr}
\eeq

Before determining the implications for the effective superpotential
of these transformation rules,
we pause in passing to record the discrete subgroup
of $U_\ssa(1) \times U_\ssr(1)$ which is anomaly-free. This
is most simply determined by requiring that the vacuum
angle, $\Theta_r$, for each of the gauge-group factors to
become shifted by an integer multiple of $2 \pi$. Keeping
in mind the relationship, \gbaredef, between the scalar part of the $S_r$
and the gauge couplings, $g_r$ and the vacuum angle, $\Theta_r$
we see that it is the field $L_r^2$ which has argument $\Theta_r$.
The anomaly-free discrete symmetry subgroup is therefore defined
by the conditions: $L^2_r \to e^{2 \pi i n_r} \; L^2_r$, where
$n_1$ and $n_2$ are integers. Requiring eqs.~\dshiftprodt\
and \dshiftanoms\ to have this effect implies the following
solutions for the allowed transformation parameters:
\label\afdsconds
\eq
\eqalign{
{\beta_\ssa \over 2 \pi} &= {\left(N_1 - \nth3 \, N_2\right) \, n_2
- \left(N_2 - \nth3 N_1\right) \, n_1 \over 2 (N_1^2 - N_2^2)} , \cr
{\beta_\ssr \over 2 \pi} &= {N_1 \, n_1
- N_2 \, n_2 \over 2 (N_1^2 - N_2^2)} . \cr}
\eeq
This same result can also be obtained by counting the
zero modes appearing in nonzero instanton amplitudes.

\vfill\eject
\subsection{Limiting Cases}

Besides being constrained by these symmetries,
the effective quantum action and Wilson action of the
model are also subject to boundary conditions, as
the parameters $s_1$, $s_2$ and $m$ take various
special values. This section outlines these boundary
conditions.

\topic{(I) $m \to \infty$}
In the limit of large $m$ the quark supermultiplets
of the microscopic theory must decouple, leaving
the theory of the pure gauge supermultiplet for
the gauge group $SU(N_1) \times SU(N_2)$, with
no matter. The superpotential for the quantum
effective action for this theory is well known,
being simply the sum of the result for each
of the separate gauge factors.

\topic{(II) $\Re s_2 \to \infty$}
When the gauge coupling of the $SU(N_2)$ factor
is taken to zero we are left with supersymmetric
QCD (SQCD), with $N_c = N_1$ colours and $N_f = N_2$
flavours. The global symmetry group (for $m=0$)
in this case is larger than for finite $s_2$
because the absence of $SU(N_2)$ gauge interactions
implies we are free to rotate the
fields $Q_{a\a}$ and $\twi{Q}^{a\a}$ independently
of one another. The flavour symmetry
therefore in this limit becomes $G_f = SU(N_2)
\times \twi{SU}(N_2) \times U_\ssa(1)
\times U_\ssb(1) \times U_\ssr(1)$.

\topic{(III) $\Re s_1 \to \infty$}
When the gauge coupling of the $SU(N_1)$ factor
is taken to zero we again have SQCD,
this time with $N_c = N_2$ colours and $N_f = N_1$
flavours. The global symmetry group (for $m=0$)
in this case is therefore $G_f = SU(N_1)
\times \twi{SU}(N_1) \times U_\ssa(1)
\times U_\ssb(1) \times U_\ssr(1)$.

In the special case $N_2 =2$ there is a still larger
global flavour symmetry because the
gauge representations ${\bf 2}$ and $\ol{\bf 2}$
are equivalent to one another. In this case
the flavour group becomes $G_f = SU(2 N_1) \times U_\ssa(1)
\times U_\ssr(1)$.

\endtopic

\subsection{The Effective Superpotential in the Confining Phase}

Consider first the confining phase ($m \ne 0$) for
which we wish to compute the superpotential for the effective
quantum action. For simplicity we consider as arguments for
this action simply the variables $M$ and $U_r$
of eq.~\conddef. Our goal is to demonstrate that the runaway
dilaton superpotential {\it is} generic for the confining
phase of the product models.

The global $U(1)$ symmetries, together with the exact
linearity requirement
that $S_r$ appear only in the term $\nth 4 \sum_r U_r S_r$,
and the quark mass appear only through the term $m M$,
determine the superpotential to have the following form:
\eq
\label\alwaysform
\eqalign{
W & =
{1 \over 32 \pi^2 } \sum_{r=1}^2
U_r \log \left( { U_r^{a_r} M^{b_r} \over
L_r^2 \; \mu_r^{3a_r + 2 b_r}} \right)
+  w\left( U_1, U_2  \right) + mM \cr
 &= \;  \frac{1}{4} U_1 S_1  +
\frac{U_1}{32\pi^2} \; \left[ a_1 \log\left(
{U_1 \over \mu_1^3} \right)
 + b_1 \log\left( {M  \over \mu_1^{2}} \right)
 \right] \cr
  & \qquad + \;
\frac{1}{4} U_2 S_2  +
\frac{U_2}{32\pi^2} \; \left[ a_2 \log\left(
{U_2 \over \mu_2^3} \right)
+ b_2 \log\left( {M \over \mu_2^{2}} \right)
 \right] + w\left( U_1, U_2
\right)+ mM . \cr}
\eeq
Here the function $w(x,y)$ which appears in eq.~\alwaysform\
is completely arbitrary, subject only to the symmetry
requirement that it be homogeneous of degree one,
\ie: $w(\lambda x,
\lambda y) \equiv \lambda w(x,y)$. The
freedom to redefine the dimensionful constants, $\mu_r$, has been used
to absorb constants which could have appeared
additively in each of the square brackets.

The constants $a_r$ and $b_r$ are determined
by requiring $U_r^{a_r} M^{b_r}/L_r^2$ to
be invariant with respect to the abelian
global symmetries. For an \Ggp\ model with
one generation of nonchiral matter in the
fundamental representation of both gauge-group
factors this implies:
\label\coupdefsfactor
\eq
a_1 = - a_2 = N_1 - N_2, \qquad  b_1 = N_2, \qquad \hbox{and} 
\qquad b_2 = N_1.
\eeq

Finally, the otherwise undetermined function, $w(x,y)$,
may be fixed by requiring $W$ to reduce to the result
for two decoupled pure gauge theories, with no matter:
\label\decoupW
\eq
W_{\rm dec} = \nth 4 \Bigl(U_1 S_1 + U_2 S_2 \Bigr)
+ {1 \over 32 \pi^2} \; \left[ N_1 U_1 \log \left(
{U_1 \over \tilde \mu_1^3} \right) + N_2 U_2 \log \left(
{U_2 \over \tilde \mu_2^3} \right) \right],
\eeq
in the decoupling limit, where the quark mass, $m$,
goes to infinity.
In this way one finds the result: $w(x,y) =
- \; {1 \over 32 \pi^2} \; \Bigl[ N_2 \, x \, \log(N_2 + N_1 y/x)
+ N_1 \, y \, \log(N_1 + N_2 x/y) \Bigr]$, allowing
the full superpotential to be written:
\eq
\label\newalwaysform
\eqalign{
W &= \;  \frac{1}{4} \Bigl(U_1 S_1  + U_2 S_2 \Bigr)
+ mM +
\frac{U_1}{32\pi^2} \; \left[ N_1 \log\left(
{U_1 \over \mu_1^3} \right)
-  N_2 \log\left( {N_2 U_1 + N_1 U_2 \over M \mu_1} \right)
 \right] \cr
  & \qquad + \;
\frac{U_2}{32\pi^2} \; \left[ N_2 \log\left(
{U_2 \over \mu_2^3} \right)
- N_1 \log\left( {N_2 U_1 + N_1 U_2\over M \mu_2} \right)
 \right] . \cr}
\eeq

It is instructive to write out the condition which is obtained
if this expression is extremized with respect to $M$.
So long as $M \ne 0$, the condition
$\partial W/\partial M = 0$ implies
\label\emresults
\eq
m M + \; {1 \over 32 \pi^2 } \; \Bigl( N_2 U_1
+ N_1 U_2 \Bigr)  = 0 ,
\eeq
which may be recognized as the Konishi anomaly. This
anomaly follows automatically from the effective superpotential
as a consequence of supersymmetry and our imposition of
the anomalous $U(1)$ symmetries.

\ref\BFQ{C.P. Burgess, A. Font and F. Quevedo, \npb{272}{86}{661}.}

Using eq.~\emresults\ to eliminate $M$
gives,\foot\rttd{Since $\ss W$ is the
superpotential for the quantum action --- as
opposed to the Wilson action --- the correct procedure
for `integrating out' fields is to remove them by solving their
extremal equations, rather than by performing their
path integral. Furthermore, for supersymmetric theories
in the low-energy limit when the fields being eliminated do not acquire
supersymmetry-breaking v.e.v.'s, this should be
done using the effective superpotential, $\ss W$, rather
than the effective scalar potential
$\ss V$ \BFQ.} {\it by construction},
the decoupled expression, eq.~\decoupW, up to an additive
constant. The scales $\tilde\mu_r$ are given by $\tilde\mu_1^3
= \mu_1^3 \; \Bigl|32\pi^2 e m/\mu_1 \Bigr|^{N_2/N_1}$,
with $\tilde\mu_2$ given by a similar expression with
$1 \leftrightarrow 2$. ($e = 2.7...$ here represents
the base of the natural logarithms.)
Varying with respect to $U_r$, for the confining phase
we therefore quite generally find the extremal value:
\label\Urreslt
\eq
\ol U_r = \left( {\tilde \mu_r^3 \over e} \right) \;
e^{- 8 \pi^2 S_r / N_r},
\eeq
and so the superpotential for the dilaton has the standard
runaway form:
\label\rwyform
\eq
W(S) = - \; {1 \over 32 \pi^2} \;
\Bigl( N_1 \, U_1 + N_2 \, U_2 \Bigr) =
- \; {1 \over 32 \pi^2 e} \;
\Bigl( N_1 \, \tilde \mu_1^3 \; e^{- 8 \pi^2 k_1 \, S/ N_1}
+ N_2 \, \tilde \mu_2^3 \; e^{- 8 \pi^2 k_2 \, S/ N_2}\Bigr),
\eeq
where eq.~\srdefs\ has been used to express $S_r$ in terms
of the dilaton $S$.

\subsection{The Wilson Superpotential for the Coulomb Phase}

We now turn our attention to the Coulomb phase, for
which $m=0$. Due to the presence of massless modes
in this phase only the superpotential for the Wilson
action is guaranteed to be local and to be a holomorphic
function of its arguments. We make some remarks
concerning the superpotential
for the effective quantum action in the Coulomb phase
in the next section.

\topic{The Choice of Variables}

Whereas the arguments of the quantum action are ours to
choose, those of the Wilson action must describe the
model's low-energy degrees of freedom. As such they can
differ for differing phases, even within a given model.

In this section we start by assuming the relevant
degrees of freedom to be similar to those which
describe the model's low-energy sector for
$N_1 > N_2$, in the limit where the
$SU(N_2)$ gauge coupling, $g_2$, is taken to zero. In
this limit we have SQCD with $N_f = N_2 <
N_c = N_1$, with the low-energy physics
described by the $N_2^2$ meson-like variables,
${\Scm_\a}^\b = Q_{a\a} \twi Q^{a\b}$. For
nonzero $g_2$ we must restrict these to be
$SU(N_2)$ invariant, and so restrict our
attention to the eigenvalues, $\lambda_p,
\; p=1,\dots,N_2$,
of the matrix ${\Scm_\a}^\b$.

Notice that, for $N_1 \ge N_2 + 2$,
these $N_2$ eigenvalues are precisely
what is required to parameterize the
model's semiclassical \Dflat\ directions even when
$g_2$ is nonzero.
For the cases $N_1 = N_2$ or $N_1 = N_2 +1$, there
are $N_2 + 1$ \Dflat\ directions, and so another invariant
is required. For example when $N_1 = N_2$ we may choose
this to be the baryonic invariants
\label\binvdef
\eq
B \equiv Q_{a_1\a_1}\cdots Q_{a_{N_2}\a_{N_2}}
\; \epsilon^{a_1 \cdots a_{N_2}} \;
\epsilon^{\a_1 \cdots \a_{N_2}}, \qquad
\twi B \equiv \twi Q^{a_1\a_1}\cdots \twi Q^{a_{N_2}\a_{N_2}}
\; \epsilon_{a_1 \cdots a_{N_2}} \;
\epsilon_{\a_1 \cdots \a_{N_2}} .
\eeq
These last two quantities
are not both independent, since
classically $B \twi B$ may be expressed as a function of the
$\lambda_p$'s.

For future use, we note in passing that one might choose
a different way to express the invariants $\lambda_p$.
This is by using quantities: $M_p \equiv \Tr \left(
\Scm^p \right) = \sum_k \lambda_k^p$, for $p = 1,\dots,N_2$.
As we shall see, there are situations for which these
variables have different physical implications, and so
care must be used when interchanging the variables
$\lambda_p$ for $M_p$. In what follows,
whenever we must choose we use
the eigenvalues $\lambda_p$, rather than the variables $M_p$.

\topic{Symmetry Constraints for the Wilson Superpotential}

Finally, notice that the dependence of the Wilson superpotential on the
invariants, $\lambda_p$, is greatly simplified by considering
its behaviour in the limit when $s_2 \to \infty$, because
of the enhanced flavour symmetry which emerges in this
limit.

Consider first the case $N_1 \ge N_2 +2$.
In this case invariance of the Wilson action with respect
to the global $U(1)$ symmetries
implies the result must have the form:
\label\wilsongen
\eq
W(S_1,S_2,\lambda_p)= \left( {L_1^2 \over
\lambda_1\cdots \lambda_{N_2}}
\right)^{1/(N_1 - N_2)}
\; \Omega (z_1,\dots,z_p),
\qquad\qquad \hbox{if}\; N_1 \ge N_2 + 2
\eeq
where $\Omega$ is an arbitrary function of the invariants
$z_p \propto L_1 L_2/\lambda_p^{(N_1 + N_2)/2}$.

Now if $L_2 \to 0$ with $L_1 L_2$ fixed, then the
promotion of the gauge $SU(N_2)$ symmetry to the
global flavour group $SU(N_2) \times \twi{SU}(N_2)$
implies the unknown function $\Omega$
can depend on the $\lambda_p$ only through the combination
$\det \Scm$. (For example, for $N_2 = 2$, $\Omega$
can depend on $\lambda_1$ and $\lambda_2$
[or: $M_1$ and $M_2$] only through the
combination $\det\Scm = \lambda_1 \lambda_2$
[or:  $\det \Scm = \hf
\left( M_1^2 - M_2 \right)$].) It follows that agreement
with this limit implies eq.~\wilsongen\ can be
sharpened to involve an unknown function of only
one variable:
\label\wilsongentwo
\eq
W(S_1,S_2,\lambda_p)= \left( {L_1^2 \over {\det \Scm}}
\right)^{1/(N_1 - N_2)}
\; \Omega (z), \qquad\qquad \hbox{if} \; N_1 \ge N_2 + 2.
\eeq
where now $\Omega$ depends only on
$z \propto L_1 L_2/(\det\Scm)^{(N_1 + N_2)/2N_2}$.

The symmetry consequences for the Wilson superpotential
are even more striking for the case where $N_1 = N_2$.
In this case the same arguments as those just given
imply that there is no superpotential at all which
is consistent with all of the symmetries of the problem.
This is because the quantum numbers for the fields
in this case ensure that any quantity which is $U_\ssa(1)$ invariant
must also be $U_\ssr(1)$ invariant. But this is inconsistent
for the superpotential, which must be invariant with
respect to $U_\ssa(1)$, but carry charge 2 with respect
to $U_\ssr(1)$. We conclude:
\label\wilsongenthree
\eq
W(S_1,S_2,\lambda_p,B,\twi B)= 0,
\qquad\qquad \hbox{if} \; N_1 = N_2.
\eeq

\vfill\eject
\topic{The Limits $g_r \to 0$}

More information about the function $\Omega(z)$
is obtained by examining the limits when either
of the gauge couplings is set to zero.
Consider first the limit where $g_2$, and so also $L_2$, vanishes,
for simplicity restricting our attention to the
case $N_1 \ge N_2 +2$. With these choices $W$ must
approach the appropriate limit, $W \propto \left( L_1^2 / \det \Scm
\right)^{1/(N_1 - N_2)}$, for SQCD with $N_f = N_2$
and $N_c = N_1$ flavours. Agreement with this limit
clearly implies $\Omega(z) \to \hbox{constant}$ as $z \to 0$.

Notice, however, that this small-$z$ behaviour
for $\Omega(z)$ implies that $W$ {\it cannot}
approach a similar finite limit which depends only
on $\det\Scm$ and $L_2$ as $L_1 \to 0$. The unique
such result consistent with the flavour
symmetries is $W \propto \left( L_2^2 / (\det\Scm)^{N_1/N_2}
\right)^{1/(N_2 - N_1)}$, which cannot be obtained if
$\Omega \sim \hbox{constant}$ for small $z$. The absence of such
a limit as $L_1 \to 0$ is just as well, however, because
the microscopic theory in this limit is SQCD with $N_f \ge N_c
+ 2$ generations, whose low-energy limit is known not
to be well-described simply by variables $\propto
Q \twi Q$ \review. We should therefore expect a transition to
another phase to qualitatively change the low-energy spectrum
when $L_1$ is sufficiently small.

\endtopic

\subsection{The Quantum Action for the Coulomb Phase}

\ref\anyhow{E. Poppitz and L. Randall, \plb{389}{96}{280},
{\tt hep-th/9608157}.}

In order to better understand the Wilson superpotential
in the Coulomb phase it is worth imagining it to have
been obtained from a quantum effective action by extremizing
with respect to the gaugino fields, $U_r$.\foot\intin{This
procedure is called `integrating in' in the second reference
of ref.~\review.} Although this procedure might seem
suspect, given the possibility for nonlocal contributions
and holomorphy anomalies, these complications have been
argued in ref.~\anyhow\ to be irrelevant under
certain circumstances.

Consider, then, the form an effective $W$ must take
consistent with ($i$) the model's global symmetries;
($ii$) the limiting behaviour for $S_r \to \infty$;
and ($iii$) its linear dependence on $S_r$. Repeating the
steps taken when analysing the confining phase gives
in this case the following result for $W$:
\eq
\label\realnewalwaysform
\eqalign{
W(S_r,U_r,\lambda_p) &= \;  \frac{1}{4} \Bigl(U_1 S_1  + U_2 S_2 \Bigr) +
\frac{U_1}{32\pi^2} \; \left[ N_1 \log\left(
{U_1 \over \mu_1^3} \right)
-  N_2 \log\left( {N_2 U_1 + N_1 U_2 \over (\det\Scm)^{1/N_2} \mu_1} \right)
 \right] \cr
  & \qquad + \;
\frac{U_2}{32\pi^2} \; \left[ N_2 \log\left(
{U_2 \over \mu_2^3} \right)
- N_1 \log\left( {N_2 U_1 + N_1 U_2\over (\det\Scm)^{1/N_2} \mu_2} \right)
 \right] , \cr}
\eeq
where $\det\Scm = \prod_p \lambda_p$.

An interesting feature of eq.~\realnewalwaysform\ is that
it is {\it not} equally valid to regard it as depending on the
two sets of variables $\lambda_p$ and $M_p$. This may be
seen by adding a quark mass term --- either $m M_1$ or
$m \left(\lambda_1 + \cdots + \lambda_{N_2}
\right)$ --- and then eliminating these variables to obtain the
superpotential for $S_r$ and $U_r$ only. For nonzero
$m$ this must reproduce the decoupled form of eq.~\decoupW.
Although eq.~\decoupW\ {\it is} reproduced if the $\lambda_p$
are used as independent variables, it is {\it not} obtained
using the $M_p$. In fact, there are no solutions
at all to $\partial W/\partial M_p = 0$, because the  mass term depends only on
$M_1$,
whereas the nonperturbative superpotential depends on all the $M_p$'s
 only through the combination $\det\Scm$.  (This situation is
not improved by adding
higher order perturbative terms to $W$.) Uncritical use of the variables
$M_p$, would lead us to conclude mistakenly that supersymmetry
is spontaneous broken.  It is possible to obtain different
physical results like this, simply by using
two different variables, because the change of variables from
$\lambda_p$ to $M_p$ is not linear and it happens that the Jacobian,
$\partial\left (\lambda_1,\cdots,\lambda_{N_2}
\right) / \partial\left (M_1,\cdots,M_{N_2} \right)$,
vanishes at the solution to the stationary condition
$\partial W/\partial \lambda_p = 0$.

If we denote by $\ol U_r$ the stationary points of
eq.~\realnewalwaysform\ with respect to variations
of the $U_r$, then:
\label\uvarresults
\eq
\eqalign{
{\Uonebar^{N_1} \over (N_2 \Uonebar + N_1 \Utwobar)^{N_2}}
&= {\kappa_1 L_1^2 \over \det\Scm} \cr
{\Utwobar^{N_2} \over (N_2 \Uonebar + N_1 \Utwobar)^{N_1}}
&= {\kappa_2 L_2^2 \over (\det\Scm)^{N_1/N_2}} , \cr}
\eeq
with the constants $\kappa_{1,2}$ defined as
$\kappa_1\equiv \mu_1^{3N_1-N_2}\;e^{N_1-N_2}$,
$\kappa_2\equiv \mu_2^{3N_2-N_1}\; e^{N_2-N_1}$.

Using these expressions the superpotential can be written as:
\label\wrest
\eq
W(S_1, S_2, \lambda_p) = W(\Uonebar, \Utwobar, \det\Scm, S_1, S_2)
=-  \; {(N_1-N_2) \over 32 \pi^2 } \;
\Bigl( \Uonebar -  \Utwobar \Bigr) ,
\eeq
so the explicit solution of $\Uonebar$ and $\Utwobar$,
using \uvarresults,
gives the desired superpotential as a function
of $L_1, L_2$ and $M$. Notice how the result would
vanish (as expected) if $N_1$ were to equal $N_2$.

It only remains to solve for $\Uonebar$ and $\Utwobar$.
To do so we start by incorporating the global $U(1)$ symmetries:
\label\symforms
\eq
\Uonebar = \left( { \kappa_1 L_1^2 \over \det\Scm} \right)^{1/(N_1 - N_2)}
f_1(z), \qquad\qquad
\Utwobar = \left( { \kappa_2 L_2^2 \over
(\det\Scm)^{N_1/N_2}} \right)^{1/(N_2 - N_1)}
f_2(z),
\eeq
where we sharpen our earlier definition, and write $z =
\sqrt{\kappa_1 \kappa_2}\;  L_1 L_2 / (\det\Scm)^{(N_1 + N_2)/2N_2}$.
Moreover, eqs.~\uvarresults\ imply  $f_2(z) = z^{2 N_1/N_2(N_1 - N_2)}
\; [f_1(z)]^{N_1^2/N_2^2}$, so it suffices to solve for $f_1(z)$.
This function is determined by eq.~\uvarresults\
as the solution to the following
algebraic equation:
\label\algebeq
\eq
X^{{N_1^2}/N_2^2} -3\lambda\;  z^{-{2/{(N_1+N_2)}}}\;
 X^{{N_1}/N_2} + N_2\, X =0,
\eeq
where
\label\ancilldefs
\eq
X(z) = N_1^{ N_2^2/(N_1^2 - N_2^2)} \;
z^{2 N_2/(N_1^2 - N_2^2)} \; f_1(z)
\qquad \hbox{and}\qquad
3\;\lambda = N_1^{-N_2/(N_1+N_2)}.
\eeq

Eq.~\algebeq\ cannot be solved in closed form for
arbitrary values of $N_1$ and $N_2$, which precludes the
explicit evaluation
of eqs.~\wrest\ in the general case. (By contrast,
it is an interesting advantage of the $U_r$-dependent
superpotential, eq.~\realnewalwaysform, that it
{\it can} be found explicitly.) We therefore defer further
perusal of the solution to the following sections,
where we focus on simple special cases. In particular,
\S4\ examines in detail the case $N_1 = 4$ and $N_2 =2$,
for which eq.~\algebeq\ is cubic, and so may
be explicitly solved.


\section{The $SU(2) \times SU(2)$ Model}

Let us now specialize to the simple case $N_1=N_2=2$, which
is also examined in ref.~\SeIn. (Our results in this section essentially
reproduce those of this reference.) In this case because the
representation $({\bf 2},{\bf 2})$ is pseudoreal, we
regard our matter content to be $Q_{ia\a} \in
({\bf 2},{\bf 2})$, where $i = 1,2$ is a flavour
index. The classical flavour symmetry of the microscopic theory
in the absence of quark masses is therefore $G_f
= SU_f(2) \times U_\ssa(1) \times U_\ssr(1)$
(with $U_\ssb(1) \subset SU_f(2)$). Anomalies
break the two $U(1)$ symmetries down to the anomaly-free
$R$-symmetry whose charge for superfields is
$\twi R = R - {2 \over 3} \; A$, and so for
which $\twi R(Q_{ia\a}) = 0$.

In this case there are $N_2+1 = 3$ \Dflat\ directions, which
we may parameterize using the symmetric matrix, $M_{ij}
= Q_{ia\a} Q_{jb\b} \, \epsilon^{ab} \, \epsilon^{\a\b}$.
The flavour symmetries imply the Wilson superpotential
only depends on these variables through the
single combination $\det M$.
There are two independent invariant quantities with respect to
the two global $U(1)$ symmetries, which we take to
be $L_1/L_2$ and $\xi \propto \det M/( L_1 \, L_2)$.

\subsection{The Confining Phase}

For nonzero quark masses, $m$, we expect a confining phase
and so compute the superpotential for the
quantum effective action. The unique such superpotential
consistent with the symmetries, linearity and which gives a
decoupled result for large $m$ is:
\label\Wfortwotwo
\eq
\eqalign{
W &= {U_1\over 32 \pi^2} \; \left[ \log \left(
{\det M \over \Lambda_1^4} \right) - 2 \log \left(
{U_1 \over U_1 + U_2} \right) \right] \cr
&\qquad\qquad + {U_2\over 32 \pi^2} \; \left[ \log \left(
{\det M \over \Lambda_2^4} \right) - 2 \log \left(
{U_2 \over U_1 + U_2} \right) \right] + \Tr\Bigl( m M
\Bigr), \cr}
\eeq
where $\Lambda_r^2\equiv \mu^2 \, L_2/\left(32 \pi^2 e^\hf \right)$,
defines the RG-invariant scale for each gauge-group factor.

If eq.~\Wfortwotwo\ is first extremized with respect to $M_{ij}$,
we find the stationary point to be
$(M^{-1})_{ij} = - 32 \pi^2 \; m_{ij}/(U_1 + U_2)$.
When this is substituted back into $W$ we
obtain the usual decoupled result, eq.~\decoupW,
with $U_r$ given by $U_r \propto \L_r^2 \; \sqrt{\det m}$.
We obtain in this way
a runaway dilaton potential, as expected.

Different information may be extracted from eq.~\Wfortwotwo\
if the $U_r$ are instead eliminated before $M_{ij}$. The
saddle point conditions $\partial W/\partial U_r = 0$
imply $U_1 / U_2  = \pm \; L_1 / L_2  = \pm \;
\L_1^2 / \L_2^2$, together with the `quantum constraint':
\label\constraintform
\eq
\det M = \Bigl( \L_1^2 \pm \L_2^2 \Bigr)^2.
\eeq
In the limit in which either $L_1$ or $L_2$ vanish,
this constraint reduces to the well-known quantum
constraint of SQCD when $N_c = N_f$, and it is precisely
what is required to ensure the matching of the $B$ and
$\twi R$ anomalies in the confining phase for vacua
having $M_{11} = M_{22} = 0$, $M_{12} \ne 0$.

\subsection{The Coulomb Phase}

Semiclassically, in the absence of quark masses the
three \Dflat\ directions do not get lifted, along
which the gauge group $SU(2) \times SU(2)$ is broken
to an unbroken, diagonal $U(1)$. Furthermore, constraint
\constraintform\ does not apply to this phase, so
the massless degrees of freedom one infers for the model
therefore comprises one $U(1)$ gauge supermultiplet plus the
three gauge-neutral matter multiplets contained in $M_{ij}$.
As is easily verified, the additional
gauge multiplet cancels the contributions of $M_{12}$
to the $B$ and $\twi R$ anomalies, thereby ensuring
these anomalies continue to match in the Coulomb phase
even though constraint \constraintform\ no longer
applies there.

We now construct the Wilson action's
superpotential and gauge coupling function for these
degrees of freedom. Although our results here reproduce
those of ref.~\SeIn, we spell them out to facilitate
our presentation of the $SU(4) \times SU(2)$ model
of the next section.

\medskip\noindent
{\sl 1. The Superpotential}

\noindent
As described in \S2\ above, because of the absence of the
quark mass matrix the Wilson superpotential for the $M_{ij}$
and the dilaton is forced to vanish by the model's
$U(1)$ flavour symmetries:
\label\noWhere
\eq
W(L_r,M_{ij}) = 0.
\eeq

\medskip\noindent
{\sl 2. The Gauge Coupling Function}

\ref\SeibWitt{N. Seiberg and E. Witten, \npb{426}{94}{19};
\npb{431}{94}{484}.}

\noindent
We now turn to the coupling function, $S_{\rm eff} \equiv
- \; {i \over 4\pi} \; \tau(S_1,S_2,M_{ij})$,
for the low-energy $U(1)$ gauge multiplet.
Here we normalize $\tau$
so that its relationship with the effective
$U(1)$ gauge coupling and $\Theta$-angle is
$\tau = {\Theta_{\rm eff} \over 2 \pi}
+ { 4 \pi i \over g^2_{\rm eff}}$.
Our construction follows that of
ref.~\SeibWitt.

To construct $\tau$, we look for a function
having the following properties:
\topic{Positivity}
Because $\tau$ appears in the gauge kinetic
terms for the $U(1)$ gauge mode of the
low-energy effective action,
its imaginary part must be positive.
\topic{Duality}
$\tau$ transforms in the standard way
under the duality transformations of the
low energy effective theory:
\label\tautrans
\eq
\tau \to {A \tau + B \over C \tau + D},
\eeq
together, possibly, with an action on the
other moduli, such as $M_{ij}$. If $A$, $B$, $C$ and $D$ are
arbitrary integers, then the duality
group is $PSL(2,Z)$. Otherwise it is a
subgroup of this group.
\topic{Global Symmetries}
Invariance of the Wilson action with respect to
the global flavour symmetries of the microscopic
model is ensured if $\tau$ depends on $L_r$ and
$M_{ij}$ only through the invariant combinations
$L_1/L_2$ and $\xi \equiv \det M/\L_1^2 \, \L_2^2$.
\topic{Singularities}
Like other terms in the Wilson action,
$\tau$ may develop singularities at points in
the moduli space where otherwise massive states
come down and become massless. Here we make the
key assumption that the singularities of $\tau$
are: ($i$) at weak coupling ($\xi \to \infty$);
and ($ii$) at the confinement points --- $\xi
= \xi_\pm \equiv (\L_1^2 \pm \L_2^2)^2/\L_1^2 \;
\L_2^2$ --- and nowhere else. (Notice $\xi_\pm$
satisfy the identity $\xi_+ - \xi_- \equiv 4$.)
The singularities at these points
are argued in ref.~\SeIn\ to be due to the masslessness
there of various monopole degrees of freedom whose
condensation is responsible for the onset of confinement.

For later purposes we remark that the gaugino condensates,
$\ol{U}_r$, are also not analytic at the singular points
$\xi = \xi_\pm$, since they are nonzero in the confining
phase, but vanish whenever $\xi \ne \xi_\pm$.
\topic{The Weak-Coupling Limit}
We require, at weak coupling, that the effective coupling,
$S_{\rm eff} = - \; i \tau / 4\pi$ approach the bare coupling,
$S_1 + S_2$, corresponding to the unbroken $U(1)$ of the
gauge group.
\topic{Nonsingular $\beta$ Function}
Finally, we require the $\beta$ function, $\beta(\tau) =
(\mu^2 \partial \tau /\partial \mu^2)_{M,L_r}$ to have
no poles for $\Im \tau \ne 0$.
\endtopic

These assumptions (which need not all be independent)
are satisfied by the geometrical solution for $\tau$
which was introduced in ref.~\SeibWitt.
This solution is obtained by taking $\tau$ to be the modulus of
the torus which is defined by a cubic curve,
\label\cubiccurve
\eq
y^2 = x^3 + a \, x^2 + b \, x + c ,
\eeq
in the two-dimensional
complex plane with parameters $a$, $b$,
and $c$ given as functions of the moduli $\xi$ and $L_1/L_2$. These
functions are chosen to ensure that this torus is
singular only at the assumed points: $\xi \to \infty$
and $\xi = \xi_\pm$.

Keeping in mind the identity $\xi_+ - \xi_- = 4$,
a choice which satisfies all of the above conditions
is:
\label\twotwoGCF
\eq
a = \hf \, \Bigl( \xi_+ + \xi_- \Bigr) - \xi = \xi_- + 2 - \xi, \qquad
b = \nth4 \, \Bigl(\xi_+ - \xi_- \Bigr) = 1, \qquad c = 0 .
\eeq
The singularities of the resulting torus are determined by
the vanishing (or divergence) of the discriminant:
\label\discriminant
\eq
\eqalign{
\Delta(a,b,c) &\equiv 4 b^3 - a^2 \, b^2 - 18 abc + 4 a^3 \, c + 27 c^3 \cr
&= (\xi - \xi_-) \; ( \xi_- + 4 - \xi). \cr}
\eeq

\ref\mfbooks{N. Koblitz, {\it Introduction to Elliptic
Curves and Modular Form, 2nd Edition}, Springer-Verlag,
New York, 1993, \bk
E.T. Whittaker and G.N. Watson, {\it A Course
of Modern Analysis}, Cambridge University Press,
Cambridge, 1940.}

Given such an elliptic curve, the gauge coupling function
may be implicitly constructed using the standard modular-invariant
function $j(\tau)$, through the relation:
\label\jtoparams
\eq
j(n \tau) = { 6912 (b - \nth3 \, a^2)^3 \over \Delta(a,b,c)}
= 256 \; {[ (\xi - \xi_- - 2)^2 - 3 ]^3 \over
(\xi - \xi_-) \, (\xi - \xi_- - 4)} \; .
\eeq
Here $n$ is to be chosen to ensure that $S_{\rm eff} \to S_1 + S_2$
in the weak-coupling (\ie\ large-$\xi$) limit.

\ref\bfnrefs{J.A. Minahan and D. Nemeschansky, \npb{468}{96}{72},
{\tt hep-th/9601059};\bk
G. Bonelli and M. Matone, \prl{76}{96}{4107}, {\tt hep-th/9602174};\bk
B. Dolan, \plb{418}{98}{107}, {\tt hep-th/9710161};\bk
J.I. Latorre and C.A. L\"utken, \plb{421}{98}{217},{\tt hep-th/9711150}.}

This choice determines the model's exact $\beta$-function,
defined by the variation of $\tau$ with $\mu$ as the
moduli $M_{ij}$ and $L_r$ are held fixed, in a similar
way as has been found for $N=2$ supersymmetric $SU(2)$
gauge theory \bfnrefs. Taking $(\mu^2 \partial/\partial
\mu^2)_{M,L_r}$ of both sides of eq.~\jtoparams, and
using $\mu^2 \partial \xi / \partial\mu^2 = -2 \,\xi$,
gives:
\label\bfnexpression
\eq
\beta(\tau) \equiv \mu^2 \left( {\partial \tau
\over \partial \mu^2 } \right)_{M,L_r} = 512 \;
{(\xi - \xi_- -2) \; [(\xi - \xi_- -2)^2 - 3]^2 \;
[2 (\xi - \xi_- - 2)^2 - 9] \over  (\xi - \xi_-)^2
\; (\xi - \xi_- - 4)^2 \; n \, j'(n\tau)} .
\eeq

We next list several of the properties of $j(\tau)$
and $j'(\tau)$, which
are useful for extracting the properties of
the functions $\tau(\xi)$ and $\beta(\tau)$.
It suffices to
specify these within a fundamental domain, $F$,
obtained by identifying points in the upper-half
$\tau$-plane under the action of $PSL(2,Z)$.
(We choose for $F$ the interior, and part of the
boundary, of the standard strip, defined
by $\Im \tau > 0$, $|\Re \tau| < \hf$
and $|\tau| > 1$.)
In fact, $j(\tau)$ furnishes a one-to-one map from
$F$ to the complex Reimann
sphere \mfbooks. The properties of interest are:
\topic{Poles and Zeroes}
The positions of all of the poles and zeroes of
$j(\tau)$ and $j'(\tau)$ are known. Neither
$j(\tau)$ nor $j'(\tau)$ have any singularities
for finite $\tau$ within $F$. $j(\tau)$ has a triple zero
at the edges of $F$, when $\tau = e^{i \pi/3}$
(plus $PSL(2,Z)$ transformations of this point),
and so $j'(\tau)$ also has a double zero here.
$j'(\tau)$ has an additional zero at $\tau = i$,
since near this point $j(\tau) = 1728 + O[(\tau - i)^2]$.
\topic{Asymptopia}
As $\tau \to i \infty$, $j(\tau)$ has the following
behaviour:
\label\qexpnofj
\eq
j(\tau) = {1 \over q} + 744 + 196884 \; q + O(q^2),
\eeq
where $q \equiv e^{2 \pi i \tau}$. It follows that
$j'(\tau) = -2 \pi i/q + \cdots$ in
the same limit.
\endtopic

Using these properties it is straightforward to verify
the following properties:
\topic{Unphysical Poles}
Notice first that the zeroes of $j'(\tau)$ in the denominator of
eq.~\bfnexpression\ are cancelled by zeroes of its
numerator, leaving a well-behaved expression as
$n\tau \to e^{i\pi/3}$ and $n\tau \to i$.
\topic{Fixing $n = 2$}
To fix $n$ we examine the weak-coupling (large-$\xi$)
limit in more detail. Combining eqs.~\jtoparams\ and
\qexpnofj\ we find $j(n\tau) = 1/q^n + \cdots
= 256 \;\xi^4 + \cdots$. This is consistent with
$S_{\rm eff} = S_1 + S_2 + \cdots$ (and so
$q \propto \xi^{-2}$) only if $n=2$. (Notice this agrees with
ref.~\SeIn, once their normalization
$\tau_{16} = \Theta_{\rm eff}/\pi + 8\pi i/g^2_{\rm eff} = 2\tau$ is
taken into account.)
\topic{The Perturbative $\beta$-function}
Given the choice $n=2$, we may read off the
weak-coupling limit of the $\beta$-function defined
by eq.~\bfnexpression. Since eq.~\jtoparams\ implies
$\xi \sim 1/\left(4 \; q^\hf \right)$ for $\tau \to i \infty$, we
see that in the same limit eq.~\bfnexpression\
becomes:
\label\wkcbfn
\eq
\beta(\tau) = - \; {2 i \over \pi} + O(q) .
\eeq
The constant term in this expression
corresponds to a one-loop result. Higher-loop
contributions to $\beta$ would be proportional to powers of
$1/\tau$ and so are seen to be zero, in agreement with
standard nonrenormalization results. The subleading terms
in eq.~\wkcbfn\ are $O(q)$ and so
express instanton contributions to
the coupling-constant running.

The one-loop contribution to
eq.~\wkcbfn\ is to be compared with the general
one-loop renormalization-group running of the
gauge couplings. The standard
one-loop expression is:
\label\genrge
\eq
{4 \pi \over g^2(\mu')} = {4 \pi \over g^2(\mu)}
+  {1\over 12 \pi} \; \Bigl[ 11 \; T(A)
- 2 \; T( \Scr_{\hf}) - T(\Scr_0) \Bigr] \; \log\left(
{\mu^{'2} \over \mu^2} \right),
\eeq
where $\Scr_\hf$ denotes the gauge representation
for the model's left-handed spin-half particles
and $\Scr_0$ is the gauge representation for
its complex scalar fields. (As before $A$ represents
the adjoint representation, as is appropriate
for the spin-one particles.) Since the scale $\mu$
of eq.~\wkcbfn\ represents a scale in the macroscopic,
low-energy theory, it plays the role taken by $\mu$
(rather than $\mu'$)
in eq.~\genrge. Specializing now to the
microscopic $SU(N_1) \times SU(N_2)$ supersymmetric
gauge theory gives:
\label\rgstandard
\eq
\mu^2 {\partial \over \partial
\mu^2} \left( {4 \pi \over g_1^2} \right)
=  {N_2 - 3 N_1 \over 4 \pi},
\qquad
\mu^2 {\partial \over \partial
\mu^2} \left( {4 \pi \over g_2^2} \right)
= {N_1 - 3 N_2 \over 4 \pi} \; .
\eeq
Adding eqs.~\rgstandard\ to one another, gives
the one-loop contribution:
\label\rgstandardtau
\eq
\beta(\tau) =  -\; {i \over 2 \pi} \; (N_1 + N_2) ,
\eeq
which clearly agrees with eq.~\wkcbfn\ once specialized
to the special case $N_1 =  N_2 = 2$.
\endtopic

\subsection{Dilaton Dependence}

The low-energy dilaton dynamics is inferred by now
re-expressing $S_1$ and $S_2$ in terms of $S$. If
$k_1 = k_2 \equiv k$, as is usually the case, this implies
$L_1 = L_2 = L \equiv \exp\left[- 4\pi^2 k S\right]$.
With this choice we see that the confining
points of the space of moduli become $\xi_- = 0$
and $\xi_+ = 4$.

\ref\susyindex{E. Witten, \npb{188}{81}{513};
\npb{202}{82}{253}.}

The vanishing of the Wilson superpotential in the Coulomb phase
for this (and any other $N_1 = N_2$) model makes the low-energy
dilaton dynamics particularly simple. All that remains is to
perform the path integration over the remaining massless degrees
of freedom. On performing these integrations, for any
globally supersymmetric model
the flat directions of the Wilson superpotential also become
flat directions of the exact quantum superpotential,
so long as the dynamics of the massless modes does
not spontaneously break supersymmetry. Since
such supersymmetry breaking is forbidden to all
orders in perturbation theory, and is protected
nonperturbatively by supersymmetric index theorems \susyindex,
it is extremely unlikely in this model.
We therefore expect the
exact superpotential for $S$ to remain precisely flat.
Unfortunately, less may be said about the shape of the dilaton superpotential
away from the flat directions. This is because this shape can
depend on  the (unknown) K\"ahler potential of the low-energy
theory, or on its (calculable) gauge coupling function.

\section{The $SU(4)\times SU(2)$ Model}

Let us now consider the case $N_1=4$, $N_2=2$,
which is the simplest example of the
class of models for which $N_1 \ge N_2 + 2$.
In this case there are $N_2 = 2$ \Dflat\ directions, which
we may parameterize using the invariant eigenvalues, $\lambda_1$
and $\lambda_2$, of the two-by-two matrix,
${\Scm_\a}^\b = Q_{a\a} \twi Q^{a\b}$.
As discussed in previous sections, the Wilson superpotential
only depends on these two variables
 through a
single combination, which we denote by $M = \left(
\lambda_1 \lambda_2 \right)^\hf$.

The analysis of the confining phase for this model
proceeds just as for the general case, as described
in \S2. We therefore focus here on the Coulomb
phase of the model.

The invariant quantity with respect to the global $U(1)$
symmetries in this case is
$z = \sqrt{\kappa_1 \kappa_2} \; L_1 \, L_2/M^{3}$,
and expressions \symforms\ and \ancilldefs\
become:
\label\spforms
\eq
\eqalign{
\Uonebar &= \left( { \sqrt\kappa_1 \; L_1 \over M} \right)
f_1(z) =  \left( {\kappa_1 L_1^2 \over 4 \sqrt{\kappa_2}
\; L_2} \right)^{1/3} \; X(z), \cr
\Utwobar &= \left( { M^2 \over \sqrt{\kappa_2} L_2} \right)
f_2(z) = \nth{4} \; \left( {\kappa_1 L_1^2
\over 4 \sqrt{\kappa_2}
\; L_2} \right)^{1/3} \; X^4(z) . \cr}
\eeq

The anomaly-free discrete discrete symmetry, eq.~\afdsconds,
is in this case the group $Z_4 \times Z_4 \subset
U_\ssa(1) \times U_\ssr(1)$, where:
\label\afdsspcase
\eq
{\beta_\ssa \over 2 \pi} = {k_\ssa \over 4},
\qquad\qquad \hbox{and} \qquad\qquad
{\beta_\ssr \over 2 \pi} = {3 k_\ssr \over 4} ,
\eeq
where $k_\ssa$ and $k_\ssr$ are integers.
The action of this discrete symmetry on the fields $U_r$,
$L_r$ and $M$ therefore is:
\label\afreeaction
\eq
U_r \to e^{i \pi k_\ssr} \; U_r, \qquad
\lambda_p \to e^{i \pi k_\ssa} \; \lambda_p ,\qquad
L_1 \to e^{i \pi (k_\ssa + 5 k_\ssr)} \; L_1 ,\qquad
L_2 \to e^{i \pi (2 k_\ssa + k_\ssr)} \; L_2 .
\eeq
Clearly vacua for which $U_r$ and $\lambda_p$ are nonzero
must come in 4-dimensional $Z_4 \times Z_4$ multiplets
which differ by an overall sign change for the $\lambda_p$
and the $U_r$.

\subsection{Monodromy}

For this example the superpotential can be
explicitly found, since the algebraic equation,
eq~\algebeq, determining $X$ either implies $X=0$
or $X$ is the solution to the cubic:
\label\cubic
\eq
X^3 -3\; \xi \;  X\;+\; 2\; =\; 0 ,
\eeq
where $\xi = \lambda \, z^{-1/3}$, and $3 \lambda = 4^{-1/3}$.
The explicit solutions of this equation can be written in terms of the
quantities
\label\lasts
\eq
T^3_{\pm}\; =\; -1\pm \sqrt{{1} - \xi^3} ,
\eeq
with the three solutions given by:
\label\threesol
\eq
X_n \equiv X(\xi;\rho_n) = \rho_n\; T_+ \; + \rho^2_n\; T_-
\eeq
with $\rho_n$ being the three roots of unity,
$\rho_n = e^{2 \pi i n/3}$, $n=0,1,2$.

We next establish that the three roots of this
cubic equation can be related to one
another by simultaneously
shifting the two vacuum angles, $\Theta_r$,
through $4 \pi$ radians (which is {\it not} a
$Z_4 \times Z_4$ transformation).
We do so by showing that such a shift can be interpreted
as a monodromy transformation about the branch point
the solutions have at $\xi^3 = 0$, or $\xi^3 = \infty$.

Consider therefore performing the simultaneous
shift $\Theta_r \to \Theta_r + 4 \pi$ for both
of the vacuum angles. Under such a shift the
argument of $L_r$ shifts by $2 \pi$, but
$\xi \propto (L_1 L_2)^{-1/3}$ acquires the
phase: $\xi \to \omega \,\xi$, where $\omega = e^{2 \pi i/3}$.
This is an invariance of the equation defining $X_n$, eq.~\cubic,
provided that $X_n$ also transforms by
$X_n \to \omega^2 X_{n'}$, since
this ensures $\omega$ cancels in both $X^3$ and $\xi X$. (Inspection
of eq.~\spforms\ shows it also leaves $U_r$ unchanged
because the transformation
$(L_1^2/L_2)^{1/3} \to \omega \,(L_1^2/L_2)^{1/3}$ cancels
the transformation of $X$.)

This shift may be regarded as a monodromy transformation since
it takes $\xi^3 \to e^{2 \pi i} \; \xi^3$, thereby circling
the branch point in the complex $\xi^3$ plane once. We now compute
its action on $X_n$. From eq.~\lasts\ we see $T^3_\pm
\left( e^{2 \pi i} \, \xi^3 \right) = T^3_\mp(\xi)$,
so if we define the branch of the cube root so that
$T_\pm\left( e^{2 \pi i} \, \xi^3 \right) =
\omega^2 \; T_\mp(\xi)$ then we find the monodromy
transformation:
\label\monodr
\eq
X(\omega \, \xi; \rho_n) = \omega^2 \; X(\xi; \rho^2_n).
\eeq

\subsection{Extremizing With Respect to $M$}

Substituting any of these solutions back into the
superpotential gives, once we use equations \wrest,
\symforms\ and \cubic:
\label\wexplicit
\eq
W_n (S_1,S_2,\lambda_1,\lambda_2)\; =  -\; {1\over{64\pi^2}}
\left({\kappa_1 L_1^2 \over{4\sqrt{\kappa_2} L_2}}
\right)^{1/3}\;
\left( 4 - X_n^3 \right)\; X_n \; ,
\eeq
for $n=0,1,2$.

Eq.~\wexplicit\ defines the superpotential as a function
of $M = \left( \lambda_1 \lambda_2\right)^\hf$.
Although $\lambda_1$ and $\lambda_2$ cannot be
separately determined, the extremal condition for $M$
is:
\label\wexplicitnom
\eq
0 = {\partial W_n \over \partial M}\;= \left( {\partial
W_n \over \partial X_n} \right) \; \left( {\partial
X_n \over \partial \xi} \right) \; \left( {\partial
\xi \over \partial M} \right) \; ,
\eeq
where
\label\factordefs
\eq
\eqalign{
\left( {\partial W_n \over \partial X_n} \right) &=
 -\; {1 \over{16\pi^2}}
\left({\kappa_1 L_1^2 \over{4\sqrt{\kappa_2} L_2}}
\right)^{1/3}\; \left( 1 - X_n^3 \right) \; , \cr
\left( {\partial X_n \over \partial \xi} \right) &=
{3 X_n^2 \over 2( X_n^3 - 1)} \; , \cr
\left( {\partial \xi \over \partial M} \right)
&= {\lambda \over (\sqrt{\kappa_1 \kappa_2} L_1 L_2)^{1/3} }\; .\cr}
\eeq

These conditions do not vanish for finite $\xi$
and $X_n$, so we are led to examine the asymptotic
behaviour for large $\xi$. This limit also permits
us to explore in detail the weak-coupling form for these
solutions. We require, then,
the large-$\xi$ limit of $T_\pm$,
which we write:
\label\bigxiTpm
\eq
T_\pm = \omega_\pm \left[ \xi^\hf \pm \; {i \over 3 \xi} + O\left(
\xi^{-{5 \over 2}} \right) \right] .
\eeq
Here $\omega_\pm^3 = \pm i$.  As may be seen from the
previous section, or by direct evaluation,
$X_n = \rho_n T_+ + \rho_n^2 T_-$ solves eq.~\cubic\ only if the phases
$\omega_\pm$ are related to one another
by $\omega_+ = i \omega_-^2$ and $\omega_- = i \sigma$,
where $\sigma$ is an arbitrary cube root of unity. (Our
phase convention for taking the cube root of $T^3_\pm$
in the previous section corresponds to the choice $\sigma =1$.)

With these phase choices in mind we obtain the following large-$\xi$
form for $\Uonebar$:
\label\smallLformone
\eq
\Uonebar = {1 \over 2 \sqrt{3}} (\omega_+ \rho_n + \omega_- \rho_n^2)
\left[ {\sqrt{\kappa_1} M L_1 \over \sqrt{\kappa_2} L_2} \right]^{1/2}
-i (- \omega_+ \rho_n + \omega_- \rho_n^2) \left[ {\sqrt{\kappa_1}
L_1 \over M} \right] + \cdots \; ,
\eeq
while $\Utwobar = \left[ (\kappa_2 L_2^2)^\hf / (\kappa_1 L_1^2) \right]
\Uonebar^4$.

\topic{The Limits $g_r \to 0$ Revisited}

We may now see explicitly how the model `chooses' to
take a simple form in the limit $L_2 \to 0$, but not
to do so for $L_1 \to 0$. That is, one expects in the limit
$L_2 \to 0$ that $\ol U_2 = 0$ and $\ol U_1$ approaches
a finite limit which is a function only of $M$ and $L_1$.
Naively, one might also expect a similar situation also as
$L_1 \to 0$, where $\ol U_1 = 0$ and $\ol U_2$ goes to
a finite limit depending only on $L_2$ and $M$.
The key observation is that, although each of the
solutions we have obtained for $\ol U_r$ indeed satisfies
{\it one} of these limits, {\it there is no one vacuum which simultaneously
satisfies both limits!}

Before exploring its implications, we first establish
the validity of this last claim. To do so notice that
for one of the three possible choices
for $\omega_\pm$ --- \ie\ for $\sigma = \rho_n^2$ --- we have
$\omega_+ \rho_n + \omega_- \rho_n^2 = 0$. With this choice the
weak-coupling limit of $\Uonebar$ is determined by the subdominant
term of eq.~\smallLformone, rather than by the leading term.
Depending on this choice, we therefore find the following two
possible weak-coupling forms for  $\Uonebar$ and $\Utwobar$:
\label\wkLformsa
\eq
\eqalign{
\Uonebar &= {1 \over 2 \sqrt{3}} (\omega_+ \rho_n + \omega_- \rho_n^2)
\left[ {\sqrt{\kappa_1} M L_1 \over \sqrt{\kappa_2} L_2} \right]^{1/2}
 + \cdots\cr
\Utwobar &= \nth{144} \; (\omega_+ \rho_n + \omega_- \rho_n^2)^4
\left[ {M^2 \over \sqrt{\kappa_2} L_2} \right] + \cdots
\qquad\qquad\hbox{if}\;(\omega_+ \rho_n + \omega_- \rho_n^2)
\ne 0 \; , \cr}
\eeq
or
\label\wkLformsb
\eq
\eqalign{
\Uonebar &= -i (- \omega_+ \rho_n + \omega_- \rho_n^2) \left[ {\sqrt{\kappa_1}
L_1 \over M} \right] + \cdots \cr
\Utwobar &= (- \omega_+ \rho_n + \omega_- \rho_n^2)^4 \left[
{\kappa_1 \sqrt{\kappa_2} L_1^2 L_2 \over M^4} \right] + \cdots
\qquad\qquad\hbox{if}\;(\omega_+ \rho_n + \omega_- \rho_n^2)
= 0 \; . \cr}
\eeq
As advertised, although eq.~\wkLformsa\ has the expected
limiting form as $L_1 \to 0$,
it predicts $\ol U_1 \to \infty$ in the
$L_2 \to 0$ limit.  Precisely the opposite
is true of eq.~\wkLformsb,
for which the $L_2 \to 0$ limit is
as expected, but where both $\ol U_1$ {\it and}
$\ol U_2$ vanish as $L_1 \to 0$.

More generally, two branches for $X_n$ have the
generic behaviour, eq.~\wkLformsa, and so
$X \sim \xi^\hf$ for large $\xi$. These
two branches are interchanged by the monodromy transformation
discussed in the previous section. For the third
branch the leading asymptotic behaviour cancels,
leaving eq.~\wkLformsb\ (or $X \sim \xi^{-1}$ for
large $\xi$), and this branch is unchanged by a
monodromy transformation.

We can now see how the model chooses eq.~\wkLformsb\
as its limiting form, thereby ensuring a simple
limit $L_2 \to 0$. The model chooses eq.~\wkLformsb\ once
the field $M$ is allowed to relax to minimize its
energy. This may be seen
by returning to our examination of
the superpotential as a function of $M$.
Inspection of eqs.~\wexplicitnom\ and
\factordefs\ shows that
the two branches for which $X \sim \xi^\hf$
for large $\xi$ do not satisfy $\partial W/\partial M
= 0$. The branch having the large-$\xi$ limit
$X \sim \xi^{-1}$ {\it does} satisfy
$\partial W/\partial M = 0$ as $\xi$, and
hence also $M$, tends to infinity, due to
the vanishing of $X$ in this limit. Being the
sole supersymmetric vacuum of the three, it is
therefore the one which is energetically preferred
once $M$ is allowed to relax.

As discussed earlier, this choice is what is
expected microscopically, since
the limiting theory as $L_2  \to 0$ is
SQCD with $N_c = 4$ colours and $N_f = 2$ flavours,
which is well described by semiclassical
\Dflat\ variables, such as the $\lambda_p$ we have used.
In the other limit, $L_1 \to 0$, however, the
microscopic theory is SQCD with $N_c = 2$ and
$N_f = 4$. But for $N_c = 2$ and $N_f = 4$ the theory is in
the `conformal window', whose low-energy limit
is believed to be controlled by a nontrivial
fixed point of the gauge coupling.
\endtopic

\subsection{The Gauge Coupling Function}

Just like the $SU(2) \times SU(2)$ model of \S3,
the $SU(4) \times SU(2)$ model under consideration
has a single unbroken $U(1)$ gauge multiplet in
the low-energy sector of its Coulomb phase.
We now compute the coupling function, $S_{\rm eff} =
- i  \tau/4\pi $ for this model. Since the
logic follows that used in \S3, we describe here
only those features which differ from this
earlier discussion.
\topic{Global Symmetries}
For the model at hand, the condition of
invariance with respect to
the global flavour symmetries of the microscopic
theory requires $\tau$ to depend only
on the single invariant quantity defined above:
$w \equiv \xi^3 \propto M^3/(L_1 L_2)$.
\topic{Singularities}
Unlike for the $SU(2) \times SU(2)$ model,
in the present case we do not have a quantum
constraint which identifies the confining phase
as a particular submanifold of the Coulomb-phase
moduli. For small $m$ the shallow directions of
the scalar potential in the confining phase are
described by the same modulus, $w$, as describes
the flat directions of the Coulomb phase.

For this model we therefore instead
identify the singular points of the function
$\tau(w)$ by permitting them only where
the gaugino condensates, $U_r$ (and hence
also the low-energy superpotential, $W(w)$) become
singular. Inspection of the explicit solution,
eqs.~\threesol\ and \lasts, shows this to occur
when $w = 0$, $w = 1$ or $w \to \infty$.
\topic{The Weak-Coupling Limit}
Identification of the unbroken $U(1)$ within
the microscopic gauge group, $SU(4) \times SU(2)$ again
implies the weak-coupling boundary condition:
$S_{\rm eff} = S_1 + S_2 + \cdots$.
\endtopic

An elliptic curve which has the required singularities,
and which satisfies all of the other requirements of \S3\
is given by eq.~\cubiccurve, with
\label\fortwoGCF
\eq
a = 4 \, w - 2, \qquad
b = 1, \qquad c = 0 ,
\eeq
for which the discriminant becomes $\Delta = 16 \, w \; (1- w)$.
The corresponding expression for the gauge coupling function,
$\tau(w)$, then is:
\label\jtoparamsfortwo
\eq
j(n \tau) = 16 \; {[ (4 w - 2)^2 - 3 ]^3 \over
w \,  (w - 1)} \; .
\eeq
Again $n = 2$ is required to ensure that $S_{\rm eff} \to S_1 + S_2$
in the large-$w$ limit.

Using $\mu^2 \partial w/\partial\mu^2 = - 6 w$,
the corresponding $\beta$-function for the model then is:
\label\bfnexpfortwo
\eq
\beta(\tau) = - 96 \;
{(2   w - 1) \; [(4w -2)^2 - 3]^2 \;
[32 w \, (w - 1) - 1] \over  w
\; (w - 1)^2 \; n \, j'(n\tau)} .
\eeq

We remark on the following properties:
\topic{Unphysical Poles}
Eq.~\bfnexpfortwo\ is well-behaved, with no
poles for $\Im \tau > 0$.
\topic{Fixing $n = 2$}
The large-$w$, large-$\Im\tau$ limit
of eq.~\jtoparamsfortwo\ states $j(n\tau) \sim 1/q^n +
\cdots = (16 \, w)^4 + \cdots$. Consistency with
$S_{\rm eff} = S_1 + S_2 + \cdots$
requires $n=2$, and $w \to 1/\left(16 \, q^\hf \right)$.
\topic{The Perturbative $\beta$-function}
With $n=2$, the weak-coupling limit of
eq.~\bfnexpfortwo\ states:
\label\wkcbfnfortwo
\eq
\beta(\tau) = - \; {3 i \over \pi} + O(q) ,
\eeq
which agrees with the perturbative nonrenormalization
theorems, as well as the one-loop beta function
of eq.~\rgstandardtau\ once this is specialized to
the case $N_1 =  4$ and $N_2 = 2$.
\endtopic

\subsection{Dilaton Dependence}

As before, we obtain the dilaton dependence of these results by
substituting into them the expression, eq.~\srdefs, for $S_r$
in terms of $S$. The usual situation where $k_1 = k_2 \equiv k$ then
implies $L_1 = L_2 \equiv L = \exp\left[- 4 \pi^2 k S \right]$.
As for the previous example, we expect general results for
global supersymmetry to
preclude spontaneous supersymmetry breaking when the
massless modes are integrated out, permitting us to analyze
the theory's flat directions using only the Wilson superpotential.
The resulting low-energy dilaton potential of this model is moderately
more complicated than for the $SU(2) \times SU(2)$ theory.

Even though the superpotential does not vanish, flat directions
along which $S$ varies are easy to find. Recall that the superpotential,
eq.~\wexplicit, has the generic form:
\label\wgenformfortwo
\eq
W(M,S) = \hbox{(constant)} \; \eta \; f(\xi),
\eeq
with $\xi \propto M \; e^{4 \pi^2 k S/3}$ and
$\eta \equiv (L_1^2/L_2)^{1/3} = e^{-4\pi^2 k S/3}$.
Here $f(\xi) = (4 - X^3) X$ does not
satisfy $f'(\xi) = 0$ for any finite $\xi$,
but $f(\xi)$ is proportional to $1/\xi$ as $\xi \to \infty$.

As discussed previously, this superpotential is extremized
by $\xi \to \infty$, for {\it any} value of $\eta$. In
our previous discussions we imagined $\xi$ being driven
to $\infty$ by relaxing $M$ with $L_1$ and $L_2$ fixed.
Now we can do so using both $M$ and $S$, so long as
the combination $\xi \to \infty$. This flat
direction is one along which $S$ is free to vary.

Notice also that the gauge coupling function, $\tau$,
depends only on $\xi$ and not separately on $S$. As
$\xi$ moves to infinity to minimize the scalar potential,
the gauge coupling function $\tau(\xi)$ is itself
driven to vanishing coupling: $\tau \to i \infty$. In
this limit the low-energy $U(1)$ gauge interactions have
no effect on the dilaton scalar potential. As in our
previous example we are led to a degenerate, supersymmetric
vacuum along which the dilaton is free to vary even
after strongly-coupled, nonabelian gauge interactions
are integrated out.

\section{Conclusions}

In this paper we have analyzed in some detail the low-energy
properties of a class of $N=1$ supersymmetric gauge theories
having gauge group $SU(N_1) \times SU(N_2)$ and matter content
$({\bf N_1},{\bf N_2}) \oplus (\ol{\bf N_1}, \ol{\bf N_1})$,
with a particular eye to the dilaton scalar potential which
these models predict. We have obtained the following results:

\topic{(1)}
We have analyzed the phase diagram of these models as functions
of the free parameters, which are the quark masses, $m$, as well
as the two gauge couplings and vacuum angles, $g_1$, $g_2$,
$\Theta_1$ and $\Theta_2$. For $m$ nonzero
we have argued the theory to be in a confining phase, for which
low-energy excitations above the confining ground state are
separated from zero energy by a nonzero gap. When $m$ is zero
there are semiclassical flat directions along which the gauge
group is generically broken to several $U(1)$ factors. We expect
a Coulomb phase to exist along these flat directions. At special
points along these flat directions it is also possible to have
larger unbroken gauge symmetries, for which other phases are
possible. When $N_1 \ge N_2 + 2$ we expect another
phase transition as the $SU(N_1)$ gauge coupling, $g_1$, is turned
off and the $SU(N_2)$ gauge coupling, $g_2$, is turned on.
This expectation is based on the qualitative change
in low-energy degrees of freedom which must happen as one
moves from supersymmetric QCD with $N_1$ colours and $N_2$
flavours to supersymmetric QCD with $N_2$ colours and $N_1$
flavours.

\topic{(2)}
We have found the explicit superpotentials, $W$, for the
quantum effective action, in the confining phase of these
models, a result which was previously unknown. In this phase
this superpotential quite generally has the form, eq.~\genformforW,
of a sum of exponentials which vanish as $\Re S \to \infty$, once
all fields but the dilaton have been eliminated.
This phase therefore always suffers from the usual runaway-dilaton
problem.

For the $SU(2) \times SU(2)$ model in particular, the model's
confinement phase is subject to a nontrivial quantum constraint.
We expect the same to be true for the $SU(N)
\times SU(N)$ models more generally.

\topic{(3)}
We have stated the symmetry conditions which constrain
the superpotential of the Wilson action for
these models. For $SU(N) \times SU(N)$ models this superpotential
must vanish identically. For other gauge groups
we have shown how this superpotential
is related to the roots of an algebraic equation, which we
cannot solve in the general case. For the particular case
of the $SU(4) \times SU(2)$ model, the algebraic equation is
cubic, and we find its solutions in some detail (in the phase
whose low-energy spectrum is described by mesonic
variables, which applies for sufficiently small $g_2$).

Although the Wilson superpotentials are in general difficult
to explicitly construct, we propose that the superpotential found
by `integrating in' the gaugino fields, $U_r$, has a simple form
for the general case.

\topic{(4)}
For the $SU(2) \times SU(2)$ and $SU(4) \times SU(2)$ models
the Coulomb phase involves a single $U(1)$ gauge multiplet, and
we exhibit the gauge coupling function for this multiplet
explictly in terms of the modulus of an elliptic curve. Our result
in the $SU(2) \times SU(2)$ case agrees with those obtained
by earlier workers. In both cases our proposed coupling functions
pass many nontrivial consistency checks, such as predicting
physically-reasonable $\beta$-functions, which are
without singularities away from $\Im \tau = 0$,
and which reproduce the known weak coupling limits.
The instanton contributions to this $\beta$-function are
absolute predictions of the proposed coupling function.

\topic{(5)}
We find the dilaton superpotential for many of these models
to have flat directions which survive the integration over
the strongly-coupled nonabelian gauge interactions. For $SU(N)
\times SU(N)$ models this is connected to the absence of a
low-energy Wilson superpotential. For the $SU(4) \times SU(2)$
model the dilaton-dependent flat direction is present even
though the Wilson superpotential does not identically vanish.

\topic{(6)}
In analyzing these models we found that  special care is
necessary when choosing how to parameterize the system's moduli.  In
particular, when the moduli correspond to \Dflat\ directions in the
semiclassical limit, there are general arguments which
permit these moduli to be parameterized by holomorphic gauge invariants.
We have found that not all choices for these invariants give the
same predictions for the low-energy physics.

In particular,
there are two natural choices for holomorphic gauge invariants
that are constructed from the `meson matrix'
${\Scm_\a}^\b = Q_{a\a} \, \twi Q^{a\b}$.  The most
widely-used invariants of this sort in the literature
are the traces: $M_p = \Tr (\Scm^p)$.  An alternative choice
instead uses the eigenvalues, $\lambda_p$, of ${\Scm_\a}^\b$.
The low-energy superpotential, $W$,  we have encountered in this
paper differ in their implications, depending on whether they
are expressed in terms of the $M_p$ or the $\lambda_p$. They
differ because the Jacobian of the transformation between
these two sets of variables is singular along the stationary
points of $W$.  We argue that it is the $\lambda_p$ which
carry the correct physical implications for the analysis of
interest in this paper.
\endtopic

\ref\givp{A. Giveon and O. Pelc, {\tt hep-th/9708168}.}
\ref\han{A. Hanany and A. Zaffaroni, {\tt hep-th/9801134}.}

Finally, it may be of interest to recover these results by using
various $D$-brane and $M$-brane configurations, such as those used in
ref.~\givp, for product group models, and those of  the related
 technique of ref.~\han.

\section{Acknowledgements}

C.B. would like to acknowledge
the Instituto de F\'isica, Universidad Nacional
Aut\'onoma de M\'exico, and the Facultat de F\'isica,
Universitat de Barcelona for their kind
hospitality during this work. Our research is
funded in part by the Natural Sciences and
Engineering Research Council of Canada (NSERC),
the McGill-UNAM exchange agreement,
CONACYT 400363-5-3275-PE and the
project DGAPA IN-103997.

\listrefs

\end